\documentclass[twoside]{IEEEtran}
\usepackage{graphicx}
\usepackage{subcaption}
 \usepackage{stfloats}
\usepackage{amsmath,amsfonts}
\usepackage{mathtools}
\usepackage{algorithmic}
\usepackage{algorithm}
\usepackage{array}
\usepackage[caption=false,font=normalsize,labelfont=sf,textfont=sf]{subfig}
\usepackage{textcomp}
\usepackage{stfloats}
\usepackage{url}
\usepackage{verbatim}
\usepackage{graphicx}
\usepackage{subcaption}
\usepackage{epsfig}

\usepackage{cite}
\usepackage{booktabs} 
\usepackage{multirow}
\usepackage{float}
\usepackage{etoolbox}
\usepackage{svg} 
\makeatletter
\patchcmd{\@makecaption}
  {\scshape}
  {}
  {}
  {}
\makeatother

\usepackage{xcolor}

\begin{document}

\title{\huge{Threat Aware Task Offloading and Caching for Secure UAV Assisted Vehicular Consumer Electronics}}

\author{Xiaoteng Yang, Sunil Prajapat and Zheng  Lin
\thanks{Corresponding author: Sunil Prajapat, Zheng  Lin.}
\thanks{Xiaoteng Yang is with the School of Telecommunications Engineering, Xidian University, Xi’an 710071, China. (email: xtengyang@stu.xidian.edu.cn).
Sunil Prajapat is with the Deptartment of Computer Engineering-AI Big Data, Marwadi University , 360006 India. (e-mail: sunilprajapat645@gmail.com) .
Zheng  Lin is with the Department of Electrical and Electronic Engineering,
University of Hong Kong, Pok Fu Lam, Hong Kong SAR, China (e-mail:
linzheng@eee.hku.hk).

}
}



\maketitle

\begin{abstract}
Vehicular consumer electronics increasingly support computation-intensive and latency-sensitive services, imposing stringent efficiency, reliability, and security requirements on vehicular edge computing (VEC) systems. In dynamic vehicular environments, inference-based information leakage and anomalous communication behaviors further threaten system performance and data privacy. To address these challenges, this paper proposes a UAV-assisted cooperative VEC architecture that integrates threat-aware task offloading with intelligent spatiotemporal caching across roadside units (RSUs) and UAV edge nodes. A security-aware uplink transmission model is developed to capture potential information leakage risks and abnormal communication patterns, enabling adaptive offloading decisions. We formulate a joint optimization problem to minimize end-to-end task execution delay while improving cache utilization under limited computing and storage resources. To efficiently solve this problem, a Threat-Aware Joint Optimization (TAGO) framework is designed by combining proximal policy optimization for adaptive task offloading and a gradient-based caching update derived from the Frank–Wolfe algorithm to capture spatiotemporal service popularity. Simulation results demonstrate that the proposed approach significantly reduces task delay and improves cache efficiency compared with several baseline strategies, showing its effectiveness for secure and efficient UAV-assisted vehicular consumer electronics systems.
\end{abstract}


\section{Introduction}

\IEEEPARstart{I}{n} intelligent vehicular consumer electronics ecosystems supported by UAV-assisted edge computing~\cite{fang2024ic3m}, the integration of AI and edge intelligence enables real-time perception, decision-making, and coordination~\cite{lin2025hasfl,sun2025intra,zhang2026memmark,wei2025optimizing,fang2024automated,qu2025mobile,yuan2024satsense,zhan2025prism,lin2026gapsl}. Despite these advancements, such systems remain vulnerable to cybersecurity threats, privacy leakage, and anomalous behaviors~\cite{duan2025llm,du2025dp,zhang2025robust,zhuang2026physically,zhu2026io}. In particular, recent advances in AI have also enabled intelligent inference, adaptive eavesdropping, and anomaly-driven attacks, allowing adversaries to exploit task offloading and data transmission processes in vehicular edge computing (VEC) systems. Traditional network architectures often lack the flexibility and resilience required to operate effectively under such adversarial conditions or variable workloads, due to limited resource elasticity, weak threat tolerance, and the absence of a unified framework that jointly considers computational efficiency and communication security~\cite{r1,r3,ra3,r4,yang2025neuroagentself}. 

With the rapid development of intelligent transportation and edge-enabled consumer electronics, the scale and complexity of vehicular edge services continue to increase. Consequently, ensuring both efficient computation and secure communication has become a critical requirement for next-generation vehicular networks. Furthermore, the scope of integrated edge intelligence extends beyond UAV assisted vehicular networks to other domains such as satellite edge computing, the Internet of Things (IoT), and large-scale cyber-physical systems, emphasizing the need for generalized, secure, and efficient optimization solutions~\cite{r5,ra5,yang2026qugridquantum}.

Generative artificial intelligence (GAI) has been increasingly introduced into vehicular consumer electronics networks, where its use is no longer limited to performance-oriented optimization~\cite{r6,r7,r9,r10}. Recent studies indicate that generative models can also support adaptive decision-making and intelligent system management under dynamic and uncertain network conditions. Prior work has explored the application of GAI to threat-related analysis, intelligent security reasoning, anomaly-tolerant task scheduling, and resource management in heterogeneous and time-varying network environments~\cite{r100}. For instance, Wu et al.~employed GAI in combination with distributed reinforcement learning to address resource allocation and channel estimation in RIS-assisted networks, resulting in adaptive scheduling behavior and improved system utility~\cite{r11}. Zhang et al.~examined a GAI-based deployment approach for service function chains in multi-domain IoT systems and observed improvements in resource utilization and service performance~\cite{r12}.

In VEC scenarios, task offloading and caching decisions are not determined solely by latency or throughput. Security-related factors must also be taken into account. As a result, traditional offloading strategies that ignore physical-layer security and data exposure risks fail to provide trustworthy vehicular communication~\cite{r13}. These limitations highlight the necessity of incorporating security-aware mechanisms into the design of offloading and caching strategies for vehicular edge systems. To address performance limitations in VEC systems, various optimization strategies have been investigated. Zhao et al.~proposed a digital twin-assisted offloading mechanism using cluster-based decision compression to reduce the action space and improve execution efficiency~\cite{r14}. Liu et al.~designed a subtask scheduling mechanism based on directed acyclic graphs and integrated distributed deep reinforcement learning to optimize task delay and energy consumption under multi-vehicle collaboration~\cite{r15}. Xiao et al.~studied multi-UAV data collection and trajectory optimization, modeling the problem as mixed integer nonlinear programming and achieving efficient control through Markov decision processes and deep reinforcement learning~\cite{r16}. Although these methods significantly improve computational efficiency and latency performance, they generally assume benign network environments and provide limited protection against emerging AI-enabled inference attacks and anomalous communication behaviors.

To enhance the computational capacity and service flexibility of vehicle-mounted systems, researchers have integrated unmanned aerial vehicles (UAVs) as auxiliary edge nodes in air--ground collaborative vehicular edge computing (VEC) architectures~\cite{ra1}. With their flexible deployment, high-altitude line-of-sight communication, and dynamic load sensing capabilities, UAVs can rapidly execute computational tasks and deliver content, significantly improving task offloading efficiency and service availability. Chen et al.~\cite{r18} proposed an energy-efficient edge cloud architecture based on a three-layer game-theoretic model. Goudarzi et al.~\cite{r19} combined software-defined networking (SDN) with UAV resources to jointly optimize UAV trajectories and task offloading using a soft actor--critic algorithm.
Zhao et al.~\cite{r20} developed a multi-UAV collaborative edge computing framework incorporating cache deployment and service scheduling.
Other studies investigated two-layer optimization frameworks~\cite{r21}, joint resource allocation schemes~\cite{r22}, Lyapunov-based online algorithms~\cite{r23}, and digital twin-assisted reinforcement learning methods~\cite{r24}, collectively enhancing computational efficiency, resource utilization, and service quality in UAV-assisted VEC systems~\cite{r25}. Recently, Zhang et al.~\cite{r27} introduced a real-time scheduling framework for executing large language model reasoning tasks on edge nodes, achieving higher reasoning throughput and more efficient resource utilization. Related autonomous-driving studies have also shown that large-language-model-enhanced trajectory planning and spatiotemporal attention mechanisms can improve intelligent mobility services in IoT-enabled vehicular environments~\cite{zhang2026largelanguage,song2025smartcity}.

Nevertheless, under AI-enabled inference and anomaly attacks, the decoupling of secure task offloading and caching remains a critical bottleneck. Most existing works optimize latency or energy efficiency in isolation, overlooking how caching decisions influence security exposure and how task placement can mitigate sensitive data retransmission. To address this challenge, this paper proposes a Threat-Aware Joint Optimization (TAGO) framework that tightly couples threat-aware task offloading, intelligent edge caching, and security-aware transmission optimization within UAV-assisted vehicular consumer electronics networks. The main contributions of this work are summarized as follows:

\begin{enumerate}
\item We develop a UAV-assisted vehicular edge computing architecture that integrates threat-aware task offloading and intelligent edge caching for vehicular consumer electronics networks. The framework explicitly considers information leakage risks and anomalous communication behaviors during task transmission, enabling secure and adaptive service provisioning in dynamic vehicular environments.

\item We formulate a multi-objective optimization problem for joint task offloading and edge caching in UAV-assisted vehicular systems. The formulation captures the trade-off between execution latency, cache efficiency, and the additional overhead introduced by encryption and privacy-preserving mechanisms.

\item We propose a Threat-Aware Joint Optimization (TAGO) framework that coordinates task offloading and caching decisions. Specifically, proximal policy optimization (PPO) is employed to learn adaptive offloading strategies, while an improved Frank--Wolfe algorithm is used to update spatiotemporal caching policies. A conditional variational autoencoder is further incorporated to enhance the robustness of threat-aware policy learning in dynamic environments.

\item Extensive simulation studies demonstrate that the proposed approach significantly reduces task response time and improves cache efficiency compared to baseline methods, while maintaining resilience across varying vehicular densities and UAV deployment scales.
\end{enumerate}

The remainder of this paper is organized as follows. Section~II describes the system model and formulates the joint optimization problem. Section~III presents the proposed algorithmic framework and optimization procedure. Section~IV outlines the experimental setup and evaluation metrics. Section~V reports and analyzes the experimental results, and Section~VI concludes the paper with directions for future research.

\section{System Model and Problem Formulation}

\begin{figure}[t]
  \centering
  \includegraphics[width=\linewidth]{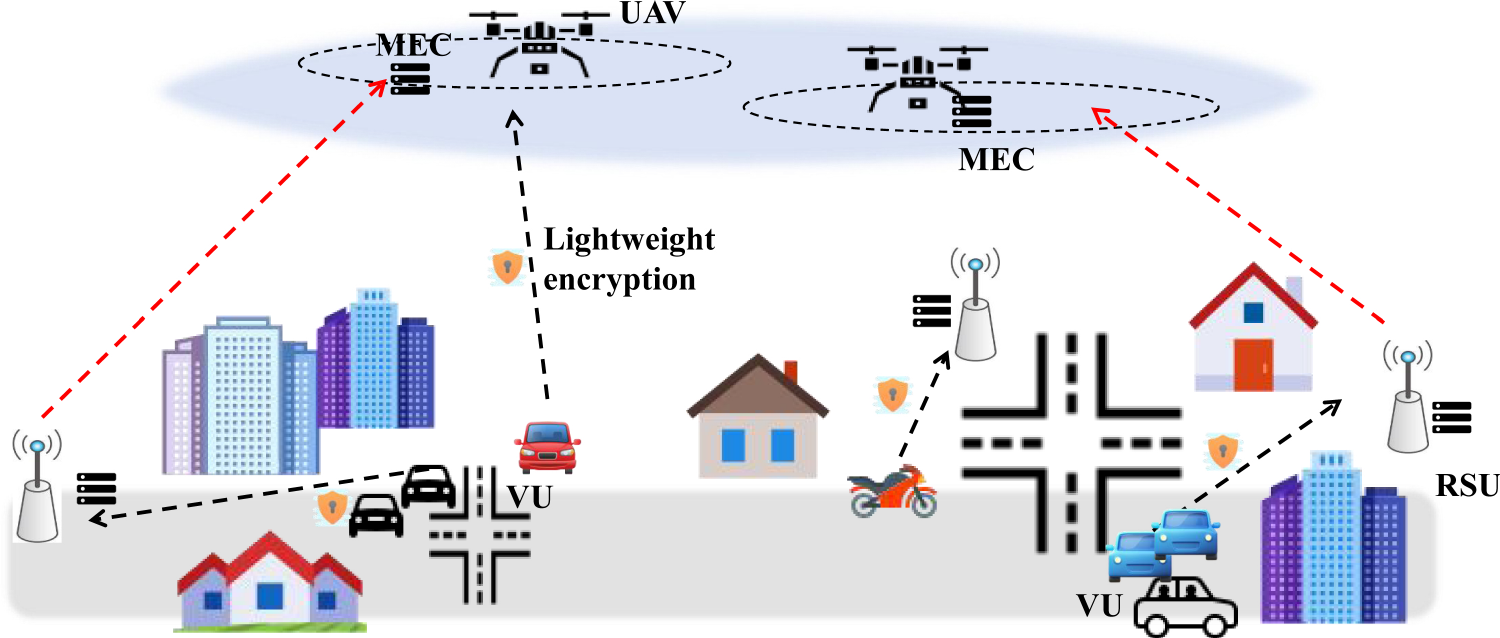}
  \caption{The proposed UAV-assisted vehicle edge computing system.}
  \label{fig_1}
\end{figure}

\subsection{System Overview}

Fig. 1 shows the architecture of a UAV-assisted vehicular consumer electronics network with NOMA-enabled transmission. The system includes a UAV $v$ with onboard computing resources, a set of vehicle users (VUs) $\mathcal{U} = \{1, 2, \dots, U\}$, and RSUs $\mathcal{S} = \{1, 2, \dots, S\}$, each providing computational and caching capabilities. During each time slot $n \in \mathcal{N} = \{1, 2, \dots, N\}$, VUs generate tasks and decide whether to execute them locally, offload to an RSU, or offload to the UAV, based on network conditions, computational load, delay requirements, and security considerations. RSUs handle the received tasks and perform data caching based on spatiotemporal access patterns. When the available resources at RSUs become insufficient, the UAV is used to supplement the computing capability and support task execution within acceptable time limits. To protect sensitive information and reduce privacy risks, tasks offloaded to RSUs or the UAV are encrypted, introducing a latency that depends on the data size. This security-aware delay is incorporated into the task scheduling model, while local execution remains unaffected. By combining encryption with anomaly-aware monitoring, the system enables adaptive, secure, and efficient offloading and caching decisions, supporting reliable operation in dynamic vehicular consumer electronics environments.
 
\subsection{Caching Model}

To support computing and caching services such as video analytics and intelligent driving for vehicle users (VUs), roadside units (RSUs) proactively cache relevant data based on task requirements. Let $\mathcal{Y} = \left \{1, 2, \dots, Y  \right \} $ denote the set of $Y$ data types requiring caching and computation. The cache state at RSU $s$ during time slot $n$ is represented as $X_s(n) =   \left \{ x_{s,1}(n), x_{s,2}(n), \dots, x_{s,Y}(n) \right \} $, where the binary variable $x_{s,y}(n) \in \left \{0,1  \right \} $ indicates whether data type $y$ is cached at RSU $s$ at time slot $n$.
The popularity of data type $y$ at RSU $s$, denoted by $p_{y,s}$, is modeled based on the Mandelbrot-Zipf (MZipf) distribution
\begin{equation}
p_{y,s} =
\frac{I_s(y)^{-z_s}}
{\sum_{y'=1}^{Y} I_s(y')^{-z_s}},
\end{equation}
where $I_s(y)$ represents the request rank of data type $y$ within the coverage area of RSU $s$ (with lower ranks indicating higher popularity), and $z_s$ is the skewness parameter controlling the popularity distribution. A larger value of $z_s$ indicates a more skewed distribution, meaning that a small number of data items dominate the request frequency, while smaller values correspond to a more uniform access pattern.

Considering the popularity of tasks, the caching policy can better predict which tasks will be selected in the future, thus improving the cache hit rate. Specifically, if RSU $s$ caches data $y$, the cache state decision variable $x_{s,y}(n)=1$ at time slot $n$, and $x_{s,y}(n)=0$ otherwise.
Meanwhile, the RSU caching strategy should be dynamically adjusted with changes in network computing resources to maximize the cache hit rate. At time slot $n$, the cache hit rate $A_s(n)$ in the RSU coverage area is defined as
\begin{equation}
   A_s(n)=
\frac{1}{|\mathcal{U}_s|}
\sum_{u\in\mathcal{U}_s} a_u(n),
\end{equation}
where $a_u(n)$ is the indicator function indicating whether the task request of VU $u$ at time slot $n$ results in a cache hit. Assuming that the data type requested by VU $u$  is $y$, when the data $y$ already exists in the cache, i.e., $x_{s,y}(n) = 1$, $a_u(n) = 1$, which indicates a hit; otherwise, $a_u(n) = 0$, which indicates a non-hit. The cache hit rate is used as a control metric for RSUs to adjust cached content in response to the access demand of VUs.

\subsection{Communication Model}

This section describes the data transmission process in the considered network. The process consists of two stages: uplink transmission from VUs to RSUs, and task forwarding from RSUs to UAVs when the available computational resources at RSUs are insufficient. A three-dimensional spatial model is adopted to characterize the wireless channels and evaluate transmission throughput, where the coordinates of network nodes are explicitly defined. The position of a VU at time slot $n$ is denoted by $(x_u(n), y_u(n), 0)$, and the associated RSU is located at $(x_s(n), y_s(n), 0)$. The horizontal Euclidean distance between the VU and RSU is thus given by

\begin{equation}
d_{u,s}(n) = \sqrt{(x_u(n) - x_s(n))^2 + (y_u(n) - y_s(n))^2}.
\end{equation}

Considering the obstacles in the environment and the fading due to environmental changes, the channel gain $g_{u,s}(n)$ from VU to RSU can be expressed as
\begin{equation}
g_{u,s}(n) = h_{u,s}(n)\,A\!\left(\frac{c}{4\pi f_c d_{u,s}(n)}\right)^{\!d_e},
\end{equation}
where $h_{u,s}(n)$ denotes the small-scale fading coefficient 
of the wireless link between VU $u$ and RSU $s$, $A$ is the 
antenna gain, $c$ is the speed of light, $f_c$ is the carrier 
frequency, and $d_e$ is the path loss exponent.

To enhance spectral efficiency, a non-orthogonal multiple access (NOMA) scheme is employed for uplink task offloading, enabling multiple VUs to simultaneously transmit data to the RSU over shared frequency resources. To manage the resulting intra-cell interference, the RSU adopts SIC, where received signals are decoded in descending order of channel gains. Specifically, the set of channel gains between all VUs and RSU $s$ at time slot $n$ is denoted as $\mathbf{G}(n) =   \left \{ g_{1,s}(n), g_{2,s}(n), \dots, g_{U,s}(n) \right \} $. Based on Shannon’s capacity theorem, the achievable uplink transmission rate of the $u$th VU to RSU $s$ is given by
\begin{equation}
R_{u,s}(n) = B_{u,s}(n) \log_2\left(1 + \frac{p_u(n) g_{u,s}(n)}{\sigma_{u,s}^2 + \sum_{i \in \mathcal{G}_u(n)} p_i(n) g_{i,s}(n)}\right)
\end{equation}
where $B_{u,s}(n)$ is the transmission channel bandwidth, $p_u(n)$ is the transmission power, $\sigma_{u,s}^2$ is the noise power, and $\mathcal{G}_u(n) = \{ i \mid g_{i,s}(n) < g_{u,s}(n) \}$ denotes the set of users whose signals are decoded after the $u$-th user in the SIC order and therefore generate interference during the decoding process.

When the computational power of the RSU cannot meet the demand, the overloaded RSU forwards the task to the UAV for processing. Assuming that the UAV is located at high altitude, its coordinates at time slot $n$ are $(x_v(n), y_v(n), z_v(n))$. At this time, the distance $d_{s,v}(n)$ between the UAV and the RSU is
\begin{equation}
    d_{s,v}(n) = \sqrt{(x_v(n) - x_s(n))^2 + (y_v(n) - y_s(n))^2 + z_v(n)^2}.
\end{equation}
Since there are fewer occlusions at high altitude, we assume that the communication between the UAV and the overloaded RSU takes place over the LoS (line-of-sight) channel. The channel transmission probability $\varrho_{s,v}(n)$ between the RSU and the UAV can be expressed as follows
\begin{equation}
    \varrho_{s,v}(n) = \frac{1}{1 + \varepsilon_1 \exp(-\varepsilon_2 (\theta_{s,v}(n) - \varepsilon_3))},
\end{equation}
where $\varepsilon_1$, $\varepsilon_2$ and $\varepsilon_3$ are environmental parameters, $\theta_{s,v}(n) = \arctan \left( \frac{z_v(n)}{\sqrt{(x_v(n) - x_s(n))^2 + (y_v(n) - y_s(n))^2}} \right)$ is the elevation angle between the RSU and the UAV. Therefore, the channel gain $g_{s,v}(n)$ of the overloaded RSU transmitting information to the UAV is
\begin{equation}
    g_{s,v}(n) = \frac{g_0 \left( \varrho_{s,v}(n) + \xi \left( 1 - \varrho_{s,v}(n) \right) \right)}{d_{s,v}^2(n)},
\end{equation}
where $g_0$ is the gain per unit distance and $\xi$ is the attenuation factor in the NLoS channel.

Ultimately, the transmission rate from the overloaded RSU to the UAV is

\begin{equation}
    R_{s,v}(n) = B_{s,v}(n) \log_2 \left( 1 + \frac{p_s(n) g_{s,v}(n)}{\sigma_{s,v}^2} \right),
\end{equation}
where $B_{s,v}(n)$ is the transmission channel bandwidth, $p_s(n)$ is the transmission power, and $\sigma_{s,v}^2$ is the noise power.

\subsection{Threat Model}

In wireless vehicular edge computing environments, task data
transmitted over open wireless channels may be exposed to
potential eavesdropping or traffic analysis attacks. An
adversary observing the communication channel may attempt
to infer sensitive task information from the transmitted
data volume and communication patterns. Let $\mathcal{L}_u(n)$ denote the information leakage level
associated with the task generated by VU $u$ at time slot $n$.
The leakage risk is assumed to increase with the amount of
transmitted data and can be modeled as $\mathcal{L}_u(n)=\eta D_u(n)$, where $D_u(n)$ is the input data size of the task and $\eta$ represents a leakage sensitivity coefficient reflecting the
inference capability of a potential adversary. Based on the leakage level, the probability that an attacker
successfully infers task-related information can be approximated as
\begin{equation}
P_u^{\text{inf}}(n)=1-\exp(-\beta \mathcal{L}_u(n)),
\end{equation}
where $\beta$ is a positive constant capturing the adversarial
inference efficiency. This formulation reflects that larger
amounts of exposed data lead to higher inference risks.

To mitigate the inference risk represented by
$P_u^{\text{inf}}(n)$, security protection mechanisms such as
encryption are applied during task transmission. These
mechanisms introduce additional computational overhead and
transmission latency. We model this overhead as a security-aware delay $t_u^{\text{sec}}(n)$, which is proportional to the leakage
level of the transmitted task $t_u^{\text{sec}}(n) = \gamma \mathcal{L}_u(n)P_u^{\text{inf}}(n),$ where $\gamma$ is a system-defined constant capturing the delay
overhead of the applied security mechanism. This formulation
reflects that higher inference risks require stronger protection
mechanisms, which introduce additional processing latency.

\subsection{Computing Model}

This section presents a detailed computational model for task processing in the proposed network. The model accounts for three types of computational execution: local processing at VUs, edge computation at RSUs, and cooperative computing assisted by UAVs. Each task generated by a VU is abstracted as a quadruple $\Phi =   \left \{D_u(n), C_u(n), O_u(n), T_u(n)  \right \} $, where $D_u(n)$ represents the input data size, $C_u(n)$ denotes the required number of CPU cycles, $O_u(n) \in \mathcal{Y}$ indicates the index of the data type to be processed, and $T_u(n)$ specifies the maximum allowable delay for task completion.

For each given VU $\alpha_{u,s}(n)=1$ indicates that the task is offloaded to the RSU for processing, and $\alpha_{u,s+1}(n)=1$ indicates that the task is executed locally at the VU. If the task is processed locally at the VU, the processing time $t_u(n)$ consumed is
\begin{equation}
    t_u(n) = \frac{C_u(n)}{Z_u},
\end{equation}
where $Z_u$ is the CPU clock speed of the VU and remains constant during the computation. If the task of the VU is offloaded to the RSU for processing, the computational delay is divided into two parts: one part is the delay of the task uploading to the neighboring RSU, and the other part is the delay of the RSU processing the task. The transmission delay of the task can be expressed as
\begin{equation}
t_{u,s}(n)=
\alpha_{u,s}(n)\frac{D_u(n)}{R_{u,s}(n)}.
\end{equation}
where $\alpha_{u,s}(n)$ is a binary offloading decision variable indicating whether the task generated by user $u$ is transmitted to RSU $s$ at time slot $n$. If $\alpha_{u,s}(n)=1$, the task is offloaded to RSU $s$, and the transmission delay depends on the achievable transmission rate $R_{u,s}(n)$; otherwise, the transmission delay is zero. The time $t_s(n)$ for the RSU to process the task is
\begin{equation}
t_s(n) = \frac{C_u(n) \cdot 
\sum_{u' \in \mathcal{U}} \alpha_{u',s}(n)}{Z_s},
\end{equation}
where $C_u(n)$ denotes the required CPU cycles of the task generated by user $u$, and $Z_s$ is the CPU clock speed of RSU $s$. The term $\sum_{u \in \mathcal{U}} \alpha_{u,s}(n)$ represents the number of tasks simultaneously assigned to RSU $s$. When $K = \sum_{u' \in \mathcal{U}} \alpha_{u',s}(n)$ tasks are 
concurrently offloaded to RSU $s$, the total computing capacity 
$Z_s$ is equally partitioned among them, such that each task 
receives an effective clock speed of $Z_s / K$ and incurs a 
processing delay of $C_u(n) \cdot K / Z_s$. If the RSU experiences heavy computational load, the remaining tasks that cannot be processed locally will be forwarded to the UAV for further processing. When the tasks are offloaded to the UAV, denoted as $\alpha_{s,v}(n)=1$, the transmission delay from RSU $s$ to the UAV can be expressed as
\begin{equation}
t_{s,v}(n) = \alpha_{s,v}(n)
\frac{D_u(n)}{R_{s,v}(n)},
\end{equation}
where $\alpha_{s,v}(n)$ is a binary decision variable 
indicating whether RSU $s$ forwards the task to the UAV, 
and $D_u(n)$ denotes the input data size of the task 
being forwarded. If $\alpha_{s,v}(n)=1$, the task is transmitted from RSU $s$ to the UAV with transmission rate $R_{s,v}(n)$; otherwise, no RSU–UAV transmission occurs. The processing delay $t_v(n)$ of the task can be expressed as
\begin{equation}
    t_v(n) = \frac{C_u(n)}{ Z_v} \cdot \max \left( 1, \sum_{u \in \mathcal{U}} \sum_{s \in \mathcal{S}} \alpha_{u,s}(n) \alpha_{s,v}(n) \right),
\end{equation}
where  $Z_v$ is the CPU clock speed of the UAV.

Based on the mutually exclusive execution locations, the total delay $t_u^{total}(n)$ for user $u$ is given by
\begin{equation}
\begin{split}
t_u^{total}(n) = \alpha_{u,s+1}(n) t_u(n) + \alpha_{u,s}(n) \Big[ t_{u,s}(n) + t_u^{\text{sec}}(n) \\
+ (1 - \alpha_{s,v}(n)) t_s(n) + \alpha_{s,v}(n) \big(t_{s,v}(n) + t_v(n)\big) \Big].
\end{split}
\end{equation}
The security-aware delay $t_u^{\text{sec}}(n)$ only applies when the task is offloaded $\alpha_{u,s}(n)=1$. This ensures that local processing is not penalized for security delay, while transmitted tasks incorporate the cost of data protection, contributing to more secure scheduling decisions.

\subsection{Problem Formulation}

In the temporal system framework $n\in \mathcal{N}$, our objective is to minimize the average task processing delay while maximizing the cache hit rate, defining the system cost $ M_n$ of the system
\begin{equation}
    M_n=\left[\lambda_1\sum_{u=1}^{U}t_{u}^{total}(n)+\sum_{s=1}^{S}\lambda_2(1-A_s(n))\right], 
\end{equation}
where $\lambda_1$ and $\lambda_2$ are positive weighting coefficients used to adjust the preferences of the different terms in the objective function to ensure that they are of the same order of magnitude in the optimization process. Thus, the joint optimization problem is formulated as follows

\allowdisplaybreaks  

\begin{align}
 & \mathbf{P}\mathbf{1} \min_{\alpha(n),x_{s,y}(n)}\frac{1}{N}\sum_{n=1}^{N}M_n \nonumber \\
 & \quad\quad\quad \mathrm{s.t.~C1:~}\alpha(n) \in \{\alpha_{u,s}(n), \alpha_{u,s+1}(n), \alpha_{s,v}(n)\}, \nonumber \\
 & \quad\quad\quad \mathrm{C2:}
    \alpha_{u,s+1}(n) + \sum_{s \in \mathcal{S}} \alpha_{u,s}(n) = 1, \quad \forall u \in \mathcal{U}, \nonumber \\ 
 & \quad\quad \quad\mathrm{C3:~}x_{s,y}(n)\in\{0,1\},\forall s\in\mathcal{S},\forall y\in\mathcal{Y}, \nonumber \\
 & \quad\quad\quad \mathrm{C4:~}\sum_{y=1}^Yl_yx_{s,y}(n)\leq L_s,\forall s\in\mathcal{S}, \nonumber \\ 
 & \quad\quad\quad \mathrm{C5:~}z_{\min}\le z_v(n)\le z_{\max}, \nonumber \\
 & \quad\quad\quad \mathrm{C6:~}\mathbf{G}(n)\in\mathbf{G}^{all}(n),\forall u\in\mathcal{U},\forall s\in\mathcal{S}, \nonumber \\
 & \quad\quad\quad \mathrm{C7:~}P_u^{\text{inf}}(n)\le P_{\max},\forall u\in\mathcal{U}, \nonumber \\
 & \quad\quad\quad \mathrm{C8:~}t_{u}^{total}(n)\leq T_u(n),\forall u\in\mathcal{U}. \label{eq:P1}
\end{align}

The optimization problem, denoted as $\mathbf{P1}$, jointly determines the task offloading and caching decisions within the UAV-assisted vehicular edge computing system. Specifically, the offloading decision variable $\alpha(n) \in   \left \{ \alpha_{u,s}(n), \alpha_{u,s+1}(n), \alpha_{s,v}(n) \right \}  $ indicates the execution location of each VU's task—either locally, at an RSU, or at a UAV. Simultaneously, the caching decision variable $x_{s,y}(n)$ represents whether data item $y$ is stored at RSU $s$, with the aim of improving data access efficiency and reducing transmission delay. Constraint $\mathrm{C1}$ and $\mathrm{C2}$ guarantee that each task is offloaded to exactly one computing entity. Constraint $\mathrm{C3}$ ensures that each RSU makes binary caching decisions per data item. Constraint $\mathrm{C4}$ imposes a storage capacity constraint such that the total size of cached data does not exceed the memory limit of the RSU. Constraint $\mathrm{C5}$ regulates the UAV’s flight altitude within a predefined operational range between $z_{\min}$ and $z_{\max}$. Constraint $\mathrm{C6}$ restricts the selected channel gain vector $\mathbf{G}(n)$ to feasible combinations in the candidate set $\mathbf{G}^{all}(n)$. Constraint $\mathrm{C7}$ ensures that the inference success probability of each VU does not exceed the predefined security threshold $P_{\max}$. Lastly, constraint $\mathrm{C8}$ ensures that the total processing delay of each task does not exceed its corresponding delay requirement $T_u(n)$.

\section{Approach Design}

Since the proposed problem $\mathbf{P1}$ is a mixed-integer nonlinear programming (MINLP) problem and belongs to the NP-hard class, directly solving it through exhaustive search is computationally infeasible for large-scale vehicular networks. To address this challenge, we propose a unified optimization framework termed Threat-Aware Generative Offloading–Caching Optimization (TAGO). The TAGO framework integrates three complementary components within a unified architecture: a reinforcement learning module for offloading optimization, a convex optimization module for cache placement, and a generative policy learning module for joint strategy refinement.

\subsection{Offloading Decision Optimization}

In this subsection, we discuss the optimization problem of offloading decisions, transform it into an MDP-based offloading problem, and propose a PPO-based offloading algorithm. The state $s(n)$ of the system in each time slice $n$ is denoted as a multidimensional state vector that covers all the network observations relevant to the offloading decision. It includes the task queue state $Q_u(n)$, which consists of the amount of data to be processed by the vehicle $u$ and the number of CPU cycles required by the task; the load state of the RSU $F_s(n)$, which consists of the current available computational resources of RSU $s$ and the total computational demand to accumulate all the queued tasks of that RSU; the channel state $G_{u,i}(n)$ and the UAV mobility state $Z_v(n)$. Thus, the system state at time slot $n$ can be defined as
\begin{equation}
    s(n)=\left \{ Q_u(n),F_s(n),G_{u,i}(n), Z_v(n)\right \}.
\end{equation}

The action space defines the offloading location selection decisions for all vehicular terminal tasks within time slot $n$. Each agent selects an offloading target from the available set comprising $S$ RSUs, the local processing unit, and $V$ UAVs. The offloading action matrix is defined as
\begin{equation}
    \alpha(n)=[\alpha_{u,j}(n)]\in\{0,1\}^{U\times(S+V+1)},
\end{equation}
where $j \in \{0,1,...,S,S+1,...,S+V\}$ denotes the local processor, RSUs, and UAVs. In order to simplify the strategy expression and action sampling in the model, the discrete action space representation of One-Hot coding is used
\begin{equation}
    a_n=\mathrm{onehot}\left(\arg\max_{j\in\{0,\dots,S+V\}}\pi_\theta(j|s(n))\right),
\end{equation}
where $\pi_\theta$ is the policy network that outputs the probability distribution of offloading actions in the current state $s_n$, considering the objective function of the offloading system, we translate delay minimization and cache hit maximization into a reward function used in PPO training. The expression for this is
\begin{equation}
    r(n)= -\lambda_1\frac{1}{U}\sum_{u=1}^{U}t_u^{total}(n)+\lambda_2\frac{1}{S}\sum_{s=1}^{S}A_s(n),
\end{equation}
where $\lambda_1$ and $\lambda_2$ are the equilibrium coefficients, and the reward function maps the two optimization objectives to the forward payoff space, which guides the policy to converge to an efficient offloading scheme. To address the high-dimensional discrete action space in task offloading, the proximal policy optimization (PPO) algorithm is employed within a reinforcement learning framework. PPO offers enhanced stability and sample efficiency, making it well-suited for complex multi-node, multi-task offloading scenarios. The approach utilizes a dual-network architecture: a policy network $\pi_\theta(a|s)$ that outputs the probability distribution over offloading targets (local, RSU, UAV), and a value network $V_\phi(s)$ that estimates the expected return of state $s$. The advantage function $\hat{A}_t$ is estimated using temporal difference methods and Generalized Advantage Estimation (GAE).

\begin{equation}
    \hat{A}_t=\delta_t+(\gamma\lambda)\delta_{t+1}+(\gamma\lambda)^2\delta_{t+2}+\cdots+(\gamma\lambda)^{T-t+1}\delta_{T-1},
\end{equation}
where $\delta_t = r_t + \gamma V_\phi(s_{t+1}) - V_\phi(s_t)$, $\gamma$ is the discount factor, and $\lambda$ is the generalized advantage estimation parameter. To balance the strategy update with the value function fit, the PPO joint multinomial loss constructs the optimisation objective and the overall loss function is defined as
\begin{equation}
    \mathcal{L}^{\mathrm{PPO}}(\theta,\phi)=\mathbb{E}_t\left[\mathcal{L}_t^{\mathrm{clip}}(\theta)-c_1\cdot\mathcal{L}_t^{\mathrm{value}}(\phi)+c_2\cdot\mathcal{L}_t^{\mathrm{ent}}(\pi_\theta)\right]
\end{equation}
where $c_1$ and $c_2$ are two hyperparameters that balance the weights of the components of the loss function. This includes the strategy loss function $\mathcal{L}_t^{\text{clip}}$ controlling the old and new strategy changes, the value function loss, and the entropy regularity term. The strategy loss function is expressed as
\begin{equation}
    \mathcal{L}_t^{\mathrm{clip}}(\theta)=\min\left(r_t(\theta)\hat{A}_t,\mathrm{clip}(r_t(\theta),1-\epsilon,1+\epsilon)\hat{A}_t\right)
\end{equation}
where $r_t(\theta) = \frac{\pi_\theta(a_t|s_t)}{\pi_{\theta_{\text{old}}}(a_t|s_t)}$ denotes the ratio of the probability of new to old strategies, and $\epsilon$ is the strategy update trimming threshold to prevent too large a strategy update. Value function loss is expressed as
\begin{equation}
    \mathcal{L}_t^\mathrm{value}(\phi)=\left(V_\phi(s_t)-R_t\right)^2,
\end{equation}
where $R_t$ denotes the real cumulative return. The entropy regularity term is denoted as
\begin{equation}
    \mathcal{L}_t^\mathrm{ent}(\pi_\theta)=-\sum_a\pi_\theta(a|s_t)\log\pi_\theta(a|s_t).
\end{equation}

During training, the policy network $\pi_\theta$ and value function network $V_\phi$ are initialized. The agent interacts with the environment by sampling actions $a_t \sim \pi_\theta(a_t|s_t)$ based on the observed state $s_t$, executes offloading decisions, receives rewards $r_t$, transitions to state $s_{t+1}$, and stores experience tuples $(s_t, a_t, r_t, s_{t+1})$. After each episode, generalized advantage estimation computes advantage values $\hat{A}_t$ and returns $R_t$. The previous policy $\pi{\theta_{\text{old}}}$ is fixed, and the current policy is updated via mini-batch training over multiple epochs using a clipped surrogate loss $\mathcal{L}_t^{\text{clip}}$ to stabilize updates. Concurrently, the value loss $\mathcal{L}_t^{\text{value}}$ and entropy regularization $\mathcal{L}_t^{\text{ent}}$ are optimized to enhance convergence and exploration. This process repeats until convergence or training completion. The trained policy network $\pi_\theta$ adaptively generates offloading decisions to jointly minimize task delay and maximize resource utilization.

\subsection{Caching Policy Optimization}

In the joint optimization problem $\mathbf{P1}$, the caching policy variable $x_{s,y}(n)$ directly influences task access delay and cache hit rate. To reduce the overall average task processing delay and improve caching efficiency, the problem is further decomposed into the subproblem $\mathbf{P2}$, which optimizes caching policies given a fixed offloading policy $\alpha(n)$. The subproblem $\mathbf{P2}$ is formulated as

\begin{equation}
    \begin{aligned}
 & \mathbf{P}\mathbf{2} \max_{x_{s,y}(n)} \sum_{s=1}^{S}A_s(n) \\
 & \quad\quad\quad \mathrm{s.t.~C3,C4}
\end{aligned} 
\end{equation}

Considering the discrete nature of the problem, coupled constraints, and the need for online updates, an enhanced Frank-Wolfe algorithm is proposed to achieve efficient and low-complexity iterative optimization. To facilitate the solution, the binary variable $x_{s,y}(n)$ is relaxed to a continuous domain, rendering the objective function convex and linear. The feasible set is defined by sparse linear inequality constraints, transforming the problem into a standard convex program

\begin{equation}
    \begin{aligned}&\max_{\mathbf{x}}\quad\mathbf{c}^{\top}\mathbf{x}\\&\mathrm{s.t.}\quad\mathbf{x}\in\mathcal{X}:=\left\{\mathbf{x}\in[0,1]^{S\times Y}\left|\sum_{y=1}^{Y}l_{y}x_{s,y}\leq L_{s},\forall s\right\}\right.\end{aligned}
\end{equation}
where $\mathbf{x}$ is the splice vector of all $x_{s,y}$ and $\mathbf{c}$ is the gradient vector with elements $c_{s,y}=\sum_{u=1}^{U}p_{u,y}\alpha_{u,s}(n)$. The standard Frank-Wolfe algorithm constructs a linear approximation model based on the current point $\mathbf{x}^{(t)}$ at each iteration and solves the subproblem over the feasible domain $\mathcal{X}$
\begin{equation}
\mathbf{d}^{(t)} = \arg\max_{x \in \mathcal{X}} 
\left\langle \nabla f(x^{(t)}), x \right\rangle.
\end{equation}
Then update the main variable
\begin{equation}
    x^{(t+1)} = x^{(t)} + \eta_t(\mathbf{d}^{(t)} - x^{(t)}), 
\quad \eta_t = \frac{2}{t+2}.
\end{equation}

To guarantee that the solution still satisfies the cache capacity limit after each round of iteration, a projection retraction operation is introduced
\begin{equation}
    x_{s,y}^{(t+1)}\leftarrow\min\left\{x_{s,y}^{(t+1)},\frac{L_s}{\sum_yl_y\cdot x_{s,y}^{(t+1)}+\epsilon}\right\},\quad\forall s,y
\end{equation}
$\epsilon > 0$ is a tiny positive number to avoid division by zero error. Since the objective of the original problem is an integer optimization problem, we adopt an integer recovery strategy. Specifically, after reaching the maximum number of iterations or convergence threshold, we recover the solution to integer form using a thresholding method for each RSU based on the final solution $\mathbf{x}^*$, i.e.
\begin{equation}
x_{s,y}^*\to\begin{cases}1&\mathrm{if}x_{s,y}^*>0.5,\\0&\mathrm{otherwise}.\end{cases}
\end{equation}

\subsection{GAI Assisted Optimization}

To further enhance the generalization capability of the system and capture the nonlinear dependencies between offloading and caching strategies, a CVAE is introduced to build a generative artificial intelligence-assisted joint optimization model. The model takes the current system state as a conditional input and is trained under the guidance of historical optimization strategies. Specifically, the offloading policy $\boldsymbol{\alpha}^{(PPO)}$ obtained through PPO and the caching strategy $\boldsymbol{x}^{(FW)}$ derived from the Frank-Wolfe method are used as prior knowledge to guide the learning process. Based on these conditions, the CVAE learns to generate a new joint strategy $(\tilde{\boldsymbol{\alpha}}, \tilde{\boldsymbol{x}})$, enabling adaptive decision-making and improved coordination between offloading and caching under dynamic network conditions.
\begin{equation}
\boldsymbol{\phi}(n)=
\begin{bmatrix}
\phi_n^{\mathrm{task}},
\phi_n^{\mathrm{network}},
\phi_n^{\mathrm{cache}}
\end{bmatrix},
\end{equation}
where $\boldsymbol{\phi}(n)$ denotes the feature representation extracted from the system state $s(n)$, including task characteristics, network conditions, and cache states. We use $(\boldsymbol{\alpha}^{(PPO)}, \boldsymbol{x}^{(FW)})$ as the optimization prior, which constitutes the CVAE conditional input $\boldsymbol{c}(n)=[s(n),\alpha(n),x(n)]$
The training objective of the CVAE is to minimize the
negative conditional variational lower bound. $\mathcal{L}_{CVAE} =
-\mathbb{E}_{q_\phi(z|\pi,c(n))}
\left[
\log p_\theta(\pi|z,c(n))
\right]
+\beta
D_{KL}
\left(
q_\phi(z|\pi,c(n))
\Vert
p(z|c(n))
\right)$ denotes the joint strategy; $q_\phi(\boldsymbol{z}|\cdot)$ is the network of encoders, which outputs the latent variable distribution; $p_\theta (\cdot|\boldsymbol{z}, \boldsymbol{c}(n)))$ is the decoder, which is used to generate the strategy; $\text{KL}[\cdot\Vert\cdot]$ is the Kullback–Leibler (KL) divergence, which is used to constrain the latent distributions; and $\beta$ is the coefficient that moderates the tradeoffs between the generation and regularity. The encoder receives the joint strategy $\boldsymbol{\pi}$ and the conditional input $\boldsymbol{c}(n))$ and learns its distribution in the latent space. Its output is the mean $\boldsymbol{\mu}_z$ and standard deviation $\boldsymbol{\sigma}_z$ of the latent variable $z$, thus defining the conditional posterior distribution
\begin{equation}
q_\phi(\boldsymbol{z}|\boldsymbol{\pi},\boldsymbol{c}(n)))=\mathcal{N}(\boldsymbol{z};\boldsymbol{\mu}_z,\mathrm{diag}(\boldsymbol{\sigma}_z^2)).
\end{equation}
We use the reparameterization trick to sample from it
\begin{equation}
    \boldsymbol{z}=\boldsymbol{\mu}_z+\boldsymbol{\sigma}_z\odot\boldsymbol{\epsilon},\quad\boldsymbol{\epsilon}\sim\mathcal{N}(0,\boldsymbol{I}).
\end{equation}
The decoder takes as input the condition $\boldsymbol{c}(n))$ and the latent variable $\boldsymbol{z}$ and generates the joint policy $\tilde{\boldsymbol{\pi}} = \left[ \tilde{\boldsymbol{\alpha}}, \tilde{\boldsymbol{x}} \right]$. The decoder is implemented by a multilayer perceptron (MLP) with the optimization objective of minimizing the difference between the generated and true policies. Decoder output
\begin{equation}
\tilde{\boldsymbol{\pi}}=p_\theta(\pi|\boldsymbol{z},\boldsymbol{c}(n))),
\end{equation}

The prior distribution $p(\boldsymbol{z}|\boldsymbol{c}(n)))$ is defined in CVAE, and a Gaussian distribution is used to predict the mean and variance of the
\begin{equation}
    p(\boldsymbol{z}|\boldsymbol{c}(n))=\mathcal{N}(\boldsymbol{\mu}_{\mathrm{prior}}(\boldsymbol{c}(n)),\boldsymbol{\sigma}_{\mathrm{prior}}^2(\boldsymbol{c}(n))).
\end{equation}

For each generated policy $\tilde{\pi}_k=[\tilde{\alpha}_k,\tilde{x}_k]$, the system utility
$M(\tilde{\pi}_k)$ is evaluated according to the objective function defined in Problem $\mathbf{P1}$, i.e.,
\begin{equation}
M(\pi)=\lambda_1\sum_{u=1}^{U}t_u^{total}(n)+\lambda_2\left(1-A_s(n)\right).
\end{equation}
The policy that minimizes the utility value is selected as the final decision. Multiple latent variables are sampled to generate candidate joint policies, and the strategy minimizing the objective function is selected as the final decision. The pseudocode for the process algorithm is shown in Algorithm 1.

\begin{algorithm}[t]
\caption{TAGO: Threat-Aware Generative Offloading–Caching Optimization}
\label{alg1}
\begin{algorithmic}[1]

\REQUIRE System state $s(n)$, policy networks $\pi_\theta,V_\phi$, cache capacity constraints
\ENSURE Optimized decisions $(\alpha^*,x^*)$

\STATE Initialize policy network $\pi_\theta$ and value network $V_\phi$
\STATE Initialize caching variables $x_{s,y}$
\STATE Initialize CVAE model $q_\psi(z|\pi,c),p_\omega(\pi|z,c(n))$

\WHILE{training not converged}

\STATE Observe current system state $s(n)$

\STATE Sample offloading action $\alpha(n) \sim \pi_\theta(\cdot|s(n))$
\STATE Execute offloading decision and obtain reward $r(n)$
\STATE Update policy and value networks via PPO objective

\STATE Compute gradient $\nabla f(x)$ based on $\alpha(n)$
\STATE Solve linear subproblem to obtain search direction $\mathbf{d}^{(t)}$
\STATE Update caching variables $x_{s,y}(n)$ via Frank–Wolfe iteration

\STATE Construct conditional context $c(n)=[s(n),\alpha(n),x_{s,y}(n)]$
\FOR{$k=1$ to $K$}
\STATE Sample latent variable $z_k \sim q_\psi(z|\pi,c(n))$
\STATE Generate candidate policy $\tilde{\pi}_k=[\tilde{\alpha}_k,\tilde{x}_k]$
\ENDFOR

\STATE Select best policy $\tilde{\pi}^*=\arg\min M(\tilde{\pi}_k)$

\IF{$M(\tilde{\pi}^*) < M(\pi)$}
\STATE Update joint policy $\pi \leftarrow \tilde{\pi}^*$
\ENDIF

\ENDWHILE

\RETURN $(\alpha^*,x^*)$

\end{algorithmic}
\end{algorithm}

\section{Experimental Settings}

A large-scale urban vehicular edge computing scenario is constructed to evaluate the proposed framework. The simulation area is defined as a $10~\text{km} \times 10~\text{km}$ urban grid, where $\left[5,10\right]$ RSUs are deployed at major road intersections to provide wireless communication and edge computing services. VUs move according to the Manhattan mobility model, which captures the grid-based mobility pattern of vehicles in urban road networks. The number of vehicles varies within the range $\left[10,30\right]$, and their speeds are randomly distributed between $\left[30,60\right]$ km/h.

To enhance communication and computation capabilities, $\left[1,10\right]$ UAVs are introduced as aerial edge nodes. The UAVs operate at altitudes between $\left[50,100\right]$ meters and dynamically adjust their positions according to the workload of nearby RSUs to provide auxiliary communication and computing services. For vehicle-to-infrastructure (V2I) communication, each RSU is allocated a bandwidth within $\left[5,20\right]$ MHz, and a NOMA scheme is adopted to improve spectrum utilization. Shadow fading is modeled as a log-normal random variable with a standard deviation between $\left[5,8\right]$ dB, while the transmission power of vehicle users is randomly selected from $\left[20,30\right]$ dBm. Task arrivals at VUs follow a Poisson process. The input data size of tasks is uniformly distributed within $\left[1,10\right]$ MB, and the required computational workload ranges from $\left[0.5,2\right] \times 10^{9}$ CPU cycles. Task popularity follows the M-Zipf distribution with skewness coefficient $z_s = 0.8$ to capture the non-uniform access characteristics of vehicular services. Each RSU caches $\left[5,20\right]$ data items, where the size of each item lies within $\left[5,10\right]$ MB. The computational frequencies of all processing nodes, including VUs, RSUs, and UAVs, vary between $\left[2,16\right]$ GHz to reflect heterogeneous computing capabilities in practical systems \cite{r15,r18,r23}.

The proposed framework is implemented in PyTorch. The PPO agent adopts a three-layer fully connected neural network with 256 neurons in each hidden layer and ReLU activation functions for both the policy and value networks. The training process uses a discount factor $\gamma = 0.99$, GAE parameter $\lambda = 0.95$, and clipping parameter $\epsilon = 0.2$. The Adam optimizer is employed with a learning rate of $3\times10^{-4}$ and a mini-batch size of 64. For caching optimization, the enhanced Frank-Wolfe algorithm runs for $T = 100$ iterations with a convergence threshold of $10^{-4}$. The CVAE module is configured with a latent dimension of 64 and KL-divergence weight $\beta = 0.5$, producing relaxed continuous representations of the offloading policy $\alpha(n)$ and caching decision $x_{s,y}(n)$ for joint strategy refinement.

To comprehensively evaluate the effectiveness of our model, we compare its performance against several baseline methods, including the following
\begin{enumerate}
    \item \emph{DO}: Delay optimization, which only considers the overall delay of the system without considering other factors.
    \item \emph{CO}: Cache optimization with the goal of maximizing the system's cache hit rate regardless of other factors.
    \item \emph{Greedy}: A greedy approach is used to assign computational tasks to UAVs, making locally optimal decisions at each step without considering long-term effects.
    \item \emph{Random}: Tasks are randomly executed at the RSU or UAV to which they are assigned and do not depend on the system state.
    \item \emph{LORA}\cite{r23}: The dynamic resource allocation problem under long-term constraints is transformed into an instantaneous decision-making problem for each time slot by means of the Lyapunov optimization framework. The objective is to minimize the long-term average task delay while satisfying the long-term constraints on energy consumption and system cost.
    \item \emph{FDLOA}\cite{r28}: Full-duplex alternating offloading and caching algorithm, which jointly optimizes task offloading and data caching decisions in full-duplex-enabled edge computing networks. By decomposing the overall problem into three subproblems, FDLOA employs an alternating optimization strategy to minimize system delay while leveraging full-duplex communication to enhance spectral efficiency.
    \item \emph{BC-A3C-GS} \cite{r29}: Bias-correction A3C with gradient sharing, which first optimizes UAV deployment to maximize vehicle coverage and energy efficiency using an adaptive population differential evolution algorithm. Then, it jointly optimizes service caching and task offloading via a DRL-based framework that enhances the traditional A3C algorithm through bias correction and global gradient sharing.
  
\end{enumerate}

\section{Experimental Results}

\subsection{Comparative analysis of task delay}

\begin{figure}[t]
  \centering
  \includegraphics[width=0.9\linewidth]{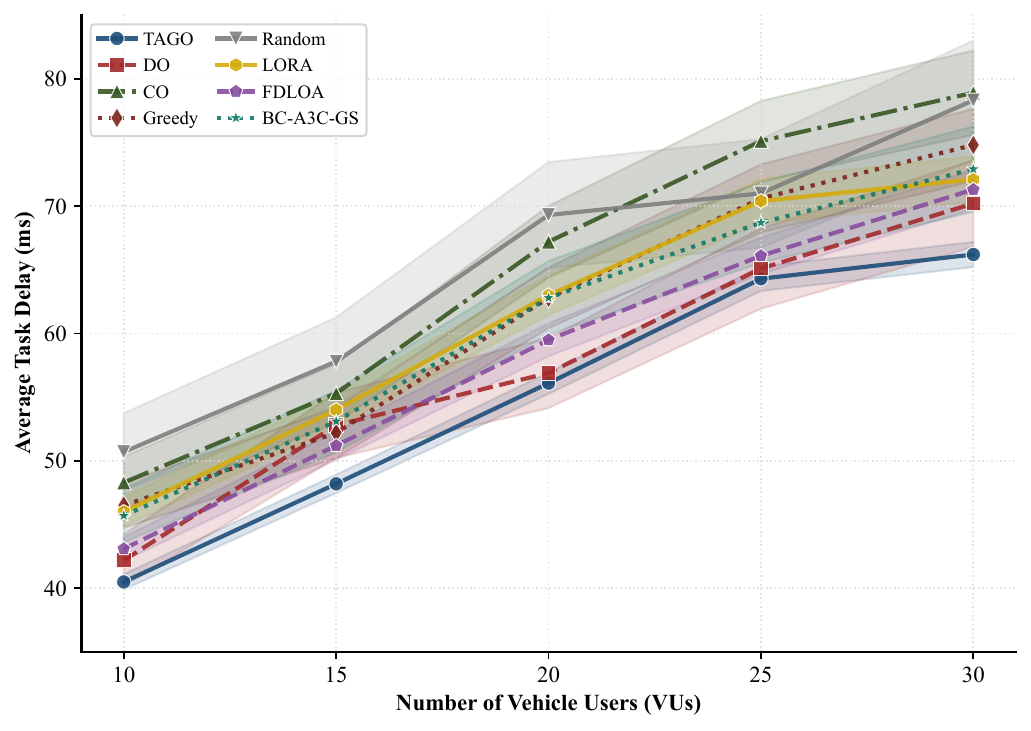}
  \caption{The impact of the number of vehicle users on the average task delay.}
  \label{fig_1}
\end{figure}

To evaluate the effectiveness of the proposed joint optimization strategy under varying system loads, we analyze the trend of average task delay as the number of vehicle users increases from 10 to 30, as shown in Fig.~2. Each curve represents the mean task delay across multiple independent runs, while the shaded regions indicate one standard deviation around the mean, capturing variability and robustness across different realizations. All methods exhibit increasing delay with higher vehicle density due to intensified competition for computational and communication resources. The proposed TAGO framework demonstrates the most moderate increase, with mean task delay rising from 40.5 to 66.2, and consistently narrow shaded regions, indicating stable performance and low variability. In contrast, the Random baseline shows the steepest increase, reaching 78.3, with wide shaded regions reflecting high fluctuation and poor adaptability under dynamic load. Traditional heuristics such as CO and Greedy, along with advanced methods including LORA, FDLOA, and BC-A3C-GS, achieve intermediate performance, with task delays 7.7\%--19.2\% higher than TAGO 
at maximum vehicle density, with CO 
showing the largest gap and FDLOA the smallest 
among this group. DO maintains closer performance to TAGO but still 
exhibits a relative gap of approximately 6\% at 
maximum vehicle density.

\begin{table*}[t]
\centering
\caption{Average Task Delay of Different Methods under Varying Vehicle Users (VUs) and UAVs.}
\label{tab:delay}
\begin{tabular}{ccccccccc}
\toprule
VUs/UAVs & Our & DO & CO & Greedy & Random & LORA & FDLOA & BC-A3C-GS \\
\midrule
\multicolumn{9}{c}{\textit{Varying Vehicle Users (VUs)}} \\
\midrule
10 & 40.5 $\downarrow$ (13.0\%) & 42.2 $\downarrow$ (9.2\%)  & 48.3 $\uparrow$ (3.9\%)  & 46.5 & 50.7 $\uparrow$ (9.0\%)  & 46.0 --           & 43.1 $\downarrow$ (7.3\%) & 45.7 --           \\
15 & 48.2 $\downarrow$ (7.7\%)  & 52.8 $\uparrow$ (1.1\%)  & 55.3 $\uparrow$ (5.9\%)  & 52.2 & 57.8 $\uparrow$ (10.7\%) & 54.0 $\uparrow$ (3.4\%)  & 51.2 $\downarrow$ (1.9\%) & 53.1 --           \\
20 & 56.1 $\downarrow$ (10.7\%) & 56.9 $\downarrow$ (9.4\%) & 67.2 $\uparrow$ (7.0\%)  & 62.8 & 69.3 $\uparrow$ (10.4\%) & 63.0 --           & 59.5 $\downarrow$ (5.3\%) & 62.8 --           \\
25 & 64.3 $\downarrow$ (8.9\%)  & 65.4 $\downarrow$ (7.4\%) & 75.1 $\uparrow$ (6.4\%)  & 70.6 & 71.0 $\uparrow$ (0.6\%)  & 70.4 --           & 66.1 $\downarrow$ (6.4\%) & 68.7 $\downarrow$ (2.7\%) \\
30 & 66.2 $\downarrow$ (11.5\%) & 70.2 $\downarrow$ (6.1\%) & 78.9 $\uparrow$ (5.5\%)  & 74.8 & 78.3 $\uparrow$ (4.7\%)  & 72.1 $\downarrow$ (3.6\%) & 71.3 $\downarrow$ (4.7\%) & 73.4 $\downarrow$ (2.5\%) \\
\midrule
\multicolumn{9}{c}{\textit{Varying Number of UAVs}} \\
\midrule
1 & 57.8 $\downarrow$ (10.4\%) & 60.2 $\downarrow$ (7.1\%) & 66.1 $\uparrow$ (2.6\%)  & 64.5 & 68.6 $\uparrow$ (6.4\%)  & 61.4 $\downarrow$ (4.8\%) & 60.3 $\downarrow$ (6.5\%) & 62.1 $\downarrow$ (3.7\%) \\
2 & 59.1 $\downarrow$ (9.8\%)  & 61.5 $\downarrow$ (5.8\%) & 67.4 $\uparrow$ (2.9\%)  & 65.3 & 69.8 $\uparrow$ (6.9\%)  & 62.6 $\downarrow$ (4.1\%) & 61.0 $\downarrow$ (6.6\%) & 63.2 $\downarrow$ (3.2\%) \\
3 & 60.4 $\downarrow$ (14.4\%) & 62.1 $\downarrow$ (12.0\%) & 73.4 $\uparrow$ (3.9\%) & 70.6 & 75.5 $\uparrow$ (6.9\%)  & 65.8 $\downarrow$ (6.8\%) & 65.2 $\downarrow$ (7.6\%) & 66.3 $\downarrow$ (6.1\%) \\
4 & 57.3 $\downarrow$ (16.8\%) & 58.9 $\downarrow$ (14.5\%) & 70.2 $\uparrow$ (1.9\%) & 68.9 & 73.8 $\uparrow$ (7.1\%)  & 61.4 $\downarrow$ (10.9\%) & 60.7 $\downarrow$ (11.9\%) & 61.5 $\downarrow$ (10.7\%) \\
5 & 54.4 $\downarrow$ (19.0\%) & 57.9 $\downarrow$ (13.8\%) & 69.5 $\uparrow$ (3.4\%) & 67.2 & 72.2 $\uparrow$ (7.4\%)  & 58.1 $\downarrow$ (13.5\%) & 57.5 $\downarrow$ (14.4\%) & 59.3 $\downarrow$ (11.7\%) \\
\bottomrule
\end{tabular}
\end{table*}

\begin{figure}[t]
  \centering
  \includegraphics[width=\linewidth]{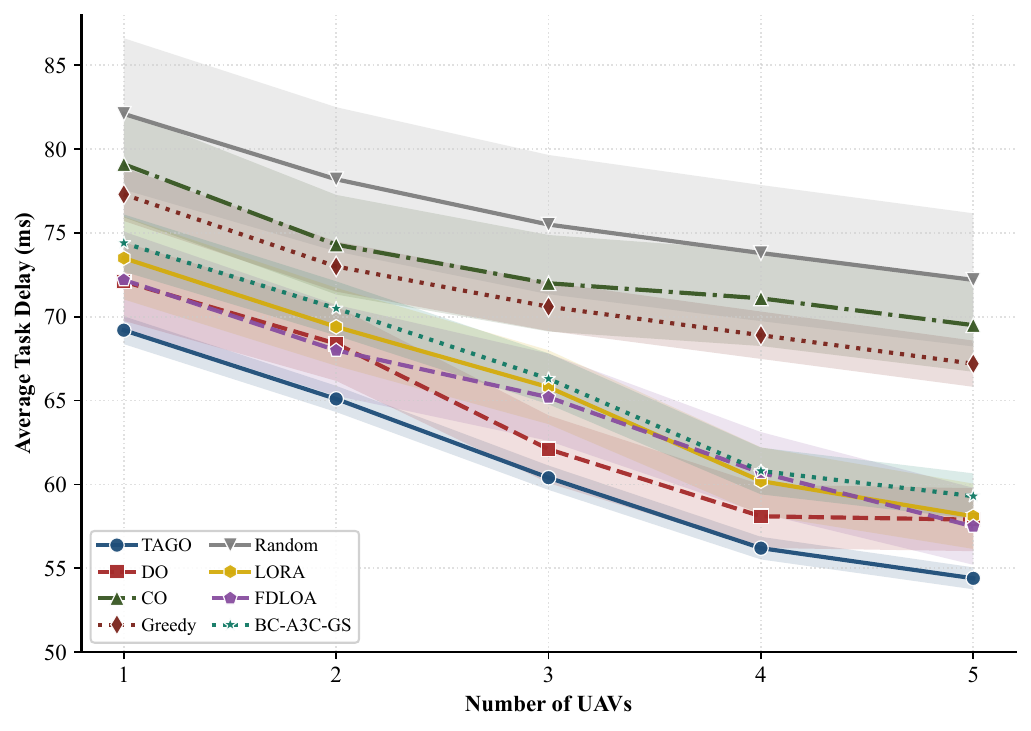}
  \caption{The impact of the number of UAVs on the average task delay.}
  \label{fig_1}
\end{figure}

\begin{figure}[t]
  \centering
  \includegraphics[width=\linewidth]{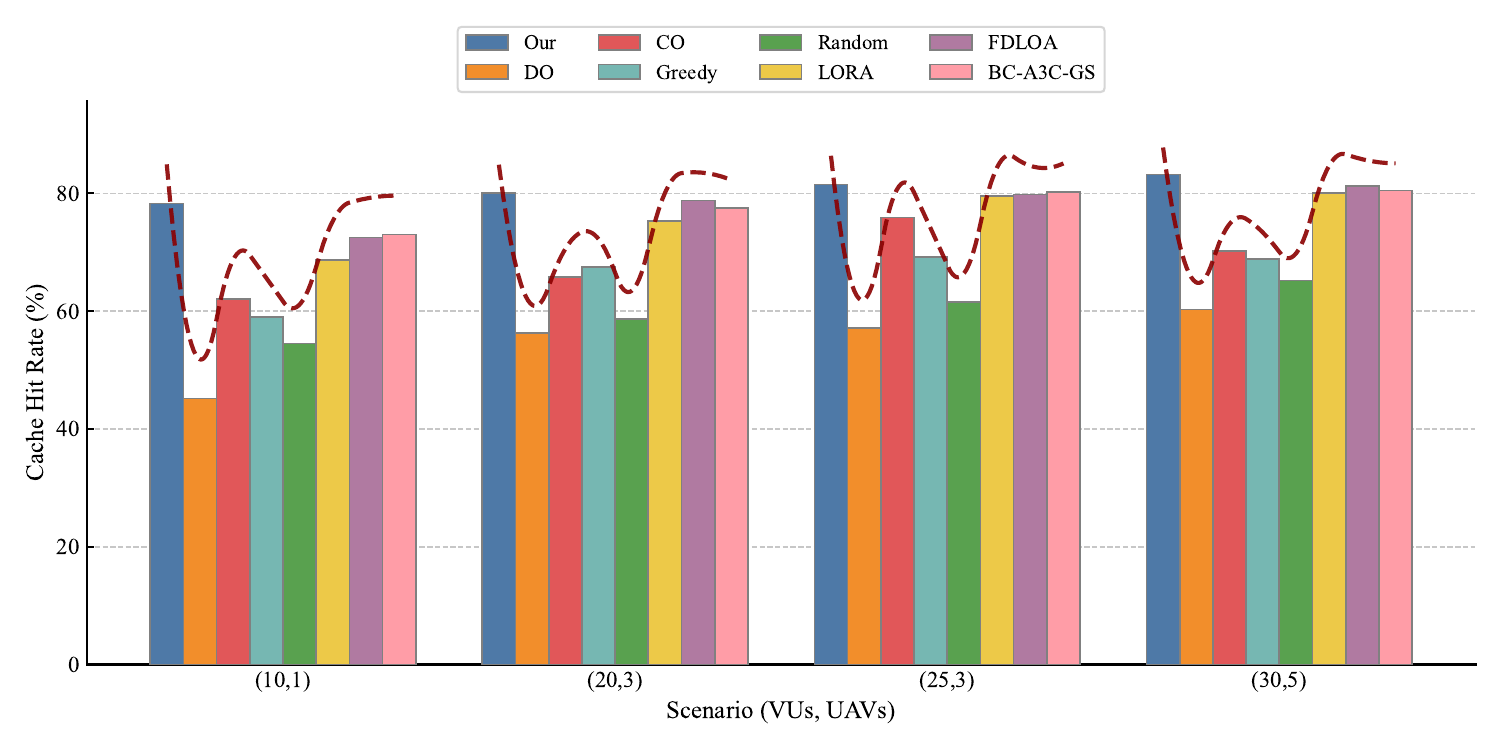}
  \caption{Comparative analysis of cache hit rates.}
  \label{fig_1}
\end{figure}

To evaluate the influence of UAV resources on task processing efficiency, we analyze the trend of average task delay as the number of UAVs increases from one to five, as shown in Fig.~3. Each curve represents the mean task delay over multiple independent experimental runs, while the shaded regions indicate one standard deviation around the mean, reflecting performance variability and algorithm robustness. The proposed TAGO framework achieves the largest 
reduction in delay, with mean values first rising 
from 57.8~ms (one UAV) to a peak of 60.4~ms 
(three UAVs) before declining to 54.4~ms (five UAVs), 
reflecting the initial coordination overhead of 
multi-UAV scheduling followed by sustained 
computational gains at larger deployment scales. In contrast, the Random and CO baselines show limited 
responsiveness to additional UAV resources, 
with delays remaining relatively flat or 
slightly increasing across UAV configurations, 
and consistently wider shaded areas indicating 
higher performance variability. LORA, FDLOA, and BC-A3C-GS achieve progressively lower 
delays with more UAVs, though they remain 3.2\%--9.8\% 
above TAGO across all configurations. The Greedy baseline 
shows a wider gap of 10.5\%--23.5\%, as local-optimal 
assignment decisions become increasingly inefficient when 
UAV load distribution shifts. DO exhibits a similar decreasing trend, yet its delay 
remains approximately 6.4\% higher than TAGO at the 
maximum UAV level, indicating that offloading decisions alone are insufficient to fully exploit the additional UAV capacity. Table~\ref{tab:delay} reports the average task delay 
of different methods under varying numbers of vehicle users and UAVs, where performance variations are shown relative to the Greedy baseline.

\subsection{Cache hit rate performance comparison}

Fig.~4 illustrates the cache hit rates achieved by different algorithms under collaborative scenarios with 10--30 vehicle users and 1--5 UAVs. Each reported result corresponds to the average performance obtained over multiple independent runs with different random seeds in order to reduce the impact of stochastic variations. For all methods, the cache hit rate increases as the system scale grows, while the increase differs across algorithms. The cache hit rates of the proposed method are 78.3\%, 80.1\%, 81.5\%, and 83.2\% in the four scenarios, respectively. BC-A3C-GS achieves cache hit rates between 73.0\% and 80.5\%, and FDLOA ranges from 72.5\% to 81.2\%. The LORA method records cache hit rates from 68.7\% to 80.4\%. DO, Greedy, and Random yield lower values, particularly in scenarios with larger numbers of vehicle users and UAVs; for instance, DO increases from 45.2\% in the $(10,1)$ scenario to 60.3\% in $(30,5)$. The distribution of cache hit rates among the 
eight methods under each scenario configuration 
is further summarized in Table~\ref{tab:cache_hit_rate}.

\subsection{Ablation experiment}

\begin{figure}[t]
  \centering
 \includegraphics[width=0.8\linewidth]{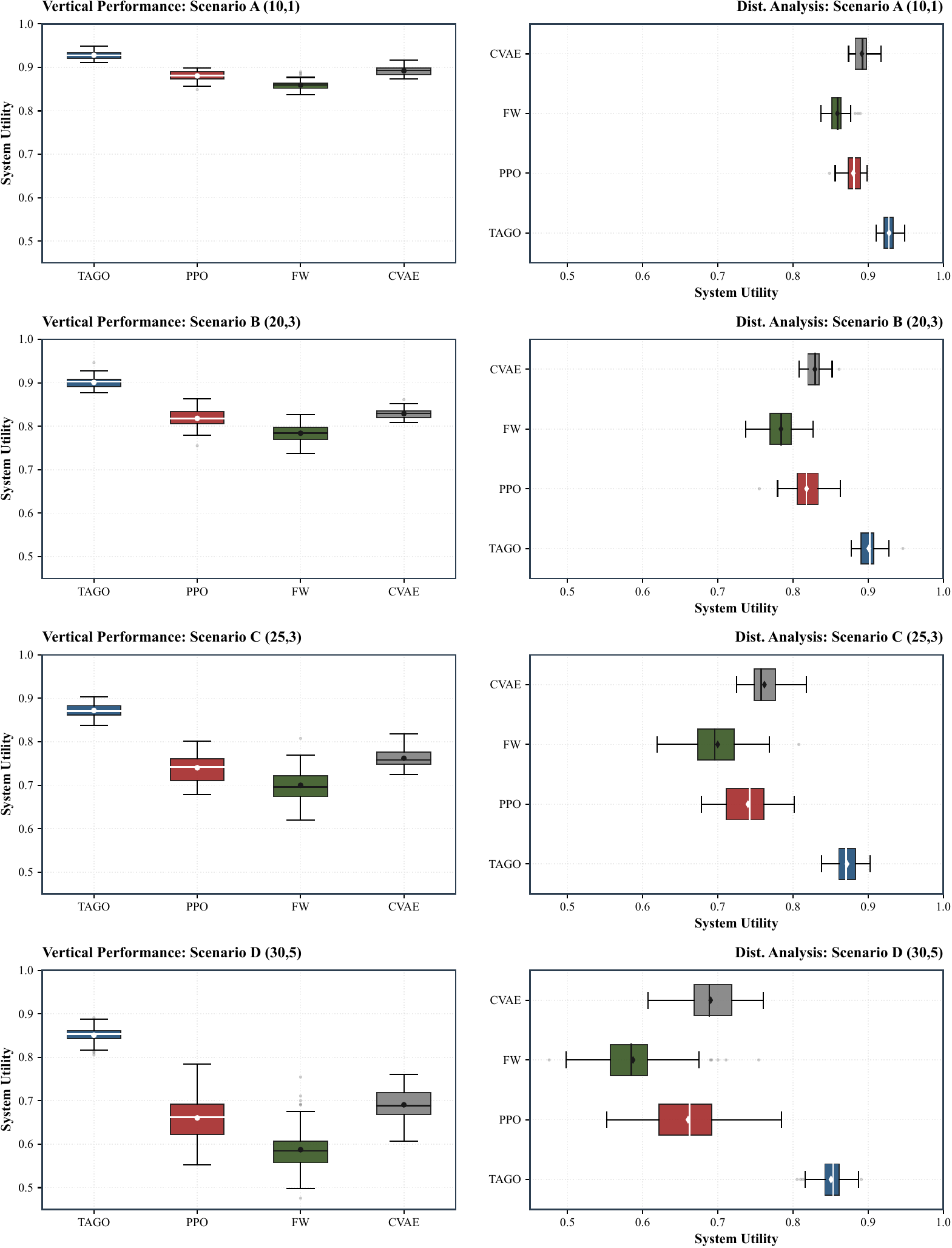}
  \caption{Comparison of results of ablation experiments.}
  \label{fig_1}
\end{figure}

To investigate the contribution of individual components, ablation experiments are conducted under four representative telematics collaboration scenarios, as shown in Fig.~5. Each subfigure visualizes the distribution of system utility using vertical and horizontal boxplots, with notches indicating median ranges and markers highlighting outliers to capture performance variability. The complete TAGO framework consistently achieves the highest system utility across all scenarios. Removing the PPO-based offloading module results in a 20$\%$–28$\%$ decrease in median utility and wider box spreads, especially under high system load. Excluding the FW-based caching optimization reduces utility by 15$\%$–20$\%$, with moderately increased variability, while disabling the CVAE-based generative module lowers median utility by around 10$\%$ and produces more outliers.

\begin{table}[t]
\centering
\caption{Cache hit rate (\%) under different configurations of VUs and UAVs, using Greedy as the baseline.}
\label{tab:cache_hit_rate}
\renewcommand{\arraystretch}{1.25}
\setlength{\tabcolsep}{5pt}
\begin{tabular}{c|cccc}
\hline
\textbf{Method} & (10,1) & (20,3) & (25,3) & (30,5) \\
\hline
Our        & \textbf{78.3} $\uparrow$32.7\% & \textbf{80.1} $\uparrow$18.7\% & \textbf{81.5} $\uparrow$17.8\% & \textbf{83.2} $\uparrow$21.0\% \\
DO         & 45.2 $\downarrow$23.4\% & 56.3 $\downarrow$16.6\% & 57.1 $\downarrow$17.5\% & 60.3 $\downarrow$12.3\% \\
CO         & 62.1 $\uparrow$5.3\% & 65.8 $\downarrow$2.5\% & \underline{75.9} $\uparrow$9.7\% & 70.2 $\uparrow$2.0\% \\
Greedy     & \textbf{59.0} & \textbf{67.5} & \textbf{69.2} & \textbf{68.8} \\
Random     & 54.5 $\downarrow$7.6\% & 58.7 $\downarrow$13.0\% & 61.5 $\downarrow$11.1\% & 65.2 $\downarrow$5.2\% \\
LORA       & \underline{68.7} $\uparrow$16.4\% & \underline{75.3} $\uparrow$11.6\% & 79.5 $\uparrow$14.9\% & \underline{80.4} $\uparrow$16.9\% \\
FDLOA      & 72.5 $\uparrow$22.9\% & 78.8 $\uparrow$16.7\% & 79.8 $\uparrow$15.3\% & 81.2 $\uparrow$18.0\% \\
BC-A3C-GS  & 73.0 $\uparrow$23.7\% & 77.5 $\uparrow$14.8\% & \underline{80.2} $\uparrow$15.9\% & 80.5 $\uparrow$17.0\% \\
\hline
\end{tabular}
\end{table}

\subsection{Model Analysis }

\begin{figure}[t]
  \centering
  \includegraphics[width=\linewidth]{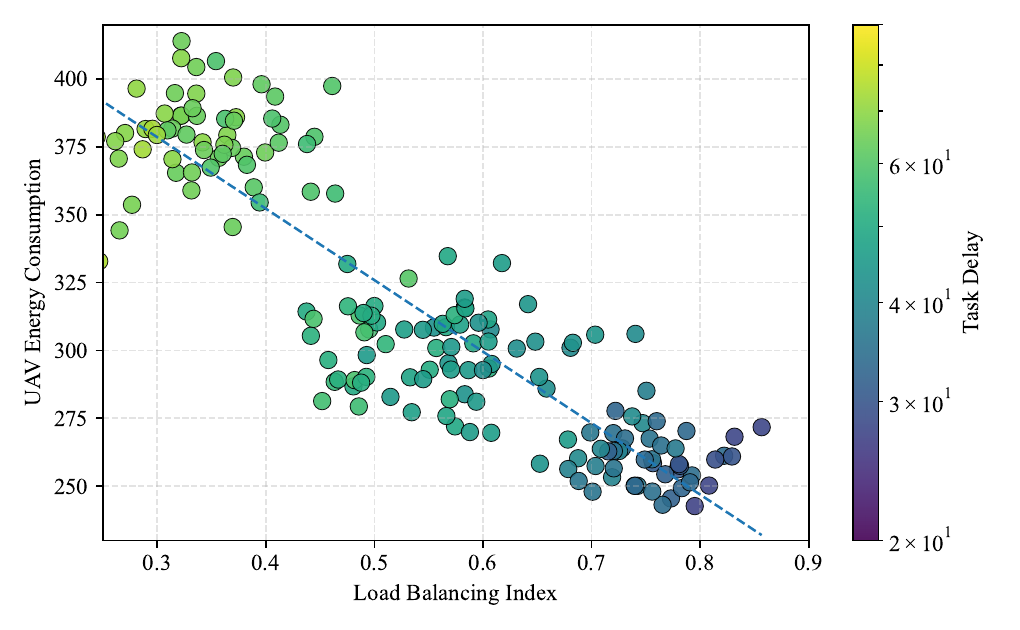}
  \caption{Energy consumption-delay trade-off analysis in UAV offloading.}
  \label{fig_1}
\end{figure}

\begin{figure}[t]
  \centering
  \includegraphics[width=0.9\linewidth]{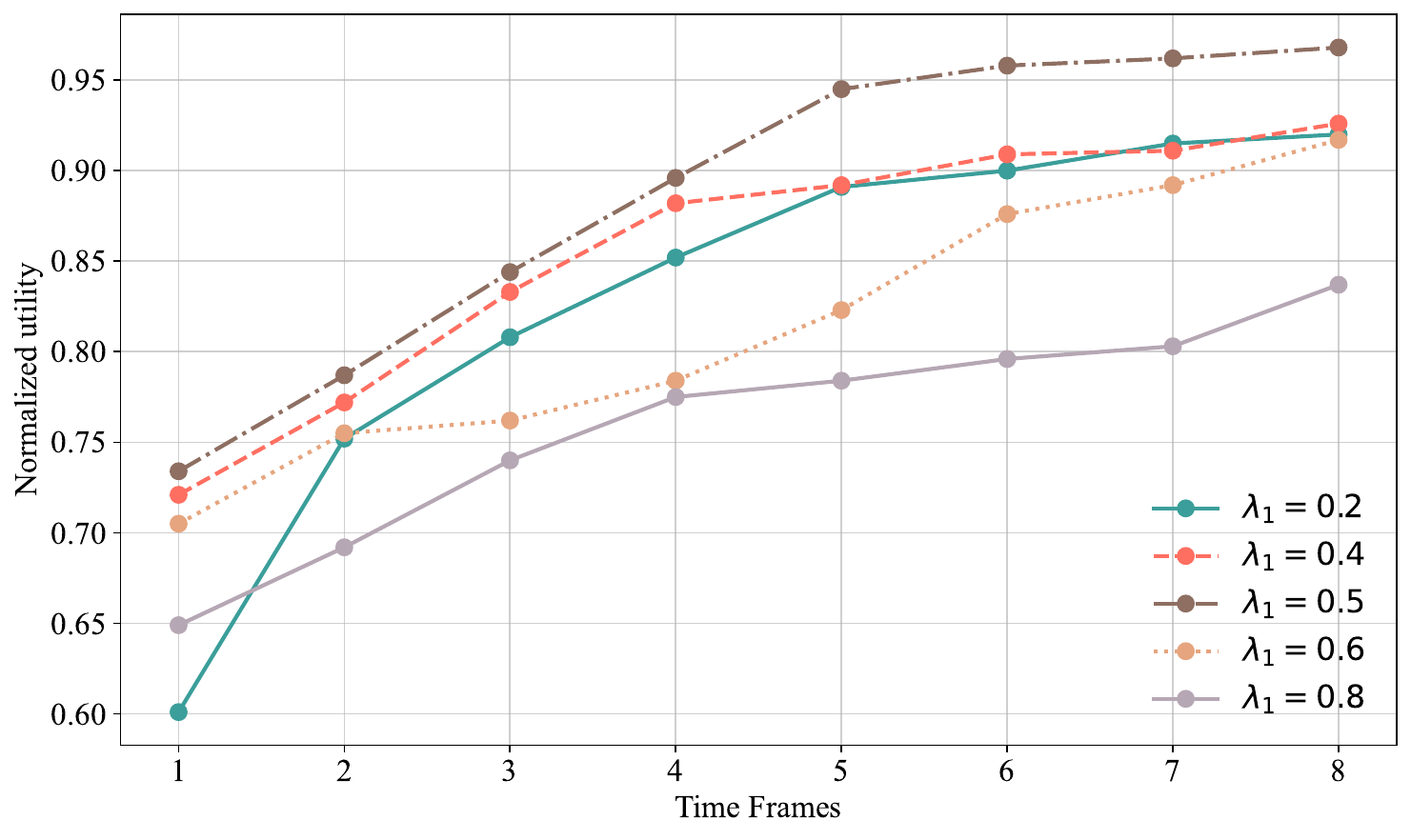}
  \caption{Model performance given different weighting coefficients $\lambda_1$.}
  \label{fig_1}
\end{figure}

\begin{figure}[t]
  \centering
  \includegraphics[width=0.9\linewidth]{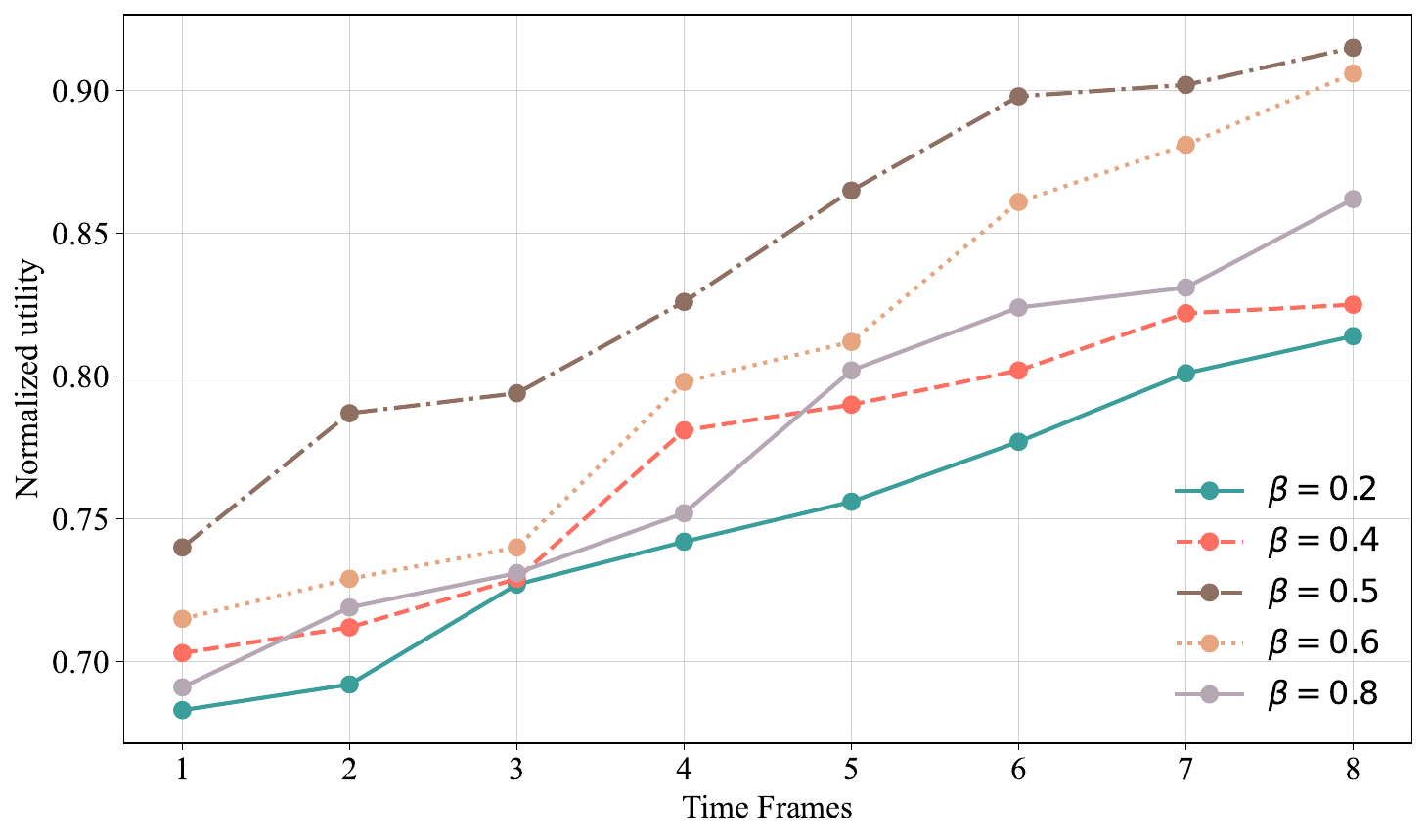}
  \caption{Model performance given different coefficients $\beta$.}
  \label{fig_1}
\end{figure}

To investigate the interaction among load distribution, UAV energy consumption, and task delay under collaborative UAV scheduling, a two-dimensional performance mapping is illustrated in Fig.~6. The horizontal axis represents the load balancing index, while the vertical axis denotes the average UAV energy consumption, with task delay encoded using a logarithmic color scale. The scatter distribution reveals a concentration 
of operating points within a load balancing index 
of 0.5--0.7 and the average UAV energy consumption remains around 260–320, within which the task delay is generally maintained below 40. In contrast, when the load distribution becomes highly uneven, certain UAV nodes experience excessive processing pressure, leading to increased energy consumption and a noticeable rise in task delay.

The impact of the weighting parameter $\lambda_1$ on system utility is illustrated in Fig.~7. A moderate $\lambda_1$ achieves faster convergence and the highest steady-state utility, yielding approximately 5\% improvement over smaller values. In contrast, overly large $\lambda_1$ values cause performance degradation exceeding 10\%, indicating diminished optimization effectiveness. The $\lambda_1$ sensitivity curves confirm a 
convergence speed versus utility trade-off, 
with moderate values yielding the best outcome.
Fig.~8 illustrates the effect of the regularization parameter $\beta$ on system performance. Moderate $\beta$ values lead to more stable utility growth and higher final performance, while insufficient or excessive regularization degrades utility by approximately 5\%--6\%. This indicates that an appropriate $\beta$ is essential for balancing latent representation flexibility and structural constraints under dynamic network conditions.

\section{Conclusion}

This paper proposed a threat-aware UAV-assisted 
framework for vehicular consumer electronics 
networks that jointly addresses secure task 
offloading and spatiotemporal caching under 
dynamic and adversarial conditions. Simulation 
results across varying vehicular densities and 
UAV deployment scales confirm that TAGO 
consistently reduces task delay and improves 
cache hit rate compared with all baselines, 
with the PPO offloading module identified as 
the dominant performance contributor through 
ablation analysis. Future work will extend the 
model to multi-UAV cooperative systems with 
joint trajectory and resource coordination, 
and will investigate more comprehensive 
adversarial models and distributed learning 
mechanisms to enhance robustness and support 
privacy-preserving optimization in large-scale 
deployments.

\bibliographystyle{IEEEtran}
\bibliography{Reference}

\begin{thebibliography}{10}
\providecommand{\url}[1]{#1}
\csname url@samestyle\endcsname
\providecommand{\newblock}{\relax}
\providecommand{\bibinfo}[2]{#2}
\providecommand{\BIBentrySTDinterwordspacing}{\spaceskip=0pt\relax}
\providecommand{\BIBentryALTinterwordstretchfactor}{4}
\providecommand{\BIBentryALTinterwordspacing}{\spaceskip=\fontdimen2\font plus
\BIBentryALTinterwordstretchfactor\fontdimen3\font minus \fontdimen4\font\relax}
\providecommand{\BIBforeignlanguage}[2]{{%
\expandafter\ifx\csname l@#1\endcsname\relax
\typeout{** WARNING: IEEEtran.bst: No hyphenation pattern has been}%
\typeout{** loaded for the language `#1'. Using the pattern for}%
\typeout{** the default language instead.}%
\else
\language=\csname l@#1\endcsname
\fi
#2}}
\providecommand{\BIBdecl}{\relax}
\BIBdecl

\bibitem{fang2024ic3m}
Z.~Fang, Z.~Lin, S.~Hu, H.~Cao, Y.~Deng, X.~Chen, and Y.~Fang, ``{IC3M: In-Car Multimodal Multi-Object Monitoring for Abnormal Status of Both Driver and Passengers},'' \emph{arXiv preprint arXiv:2410.02592}, 2024.

\bibitem{lin2025hasfl}
Z.~Lin, Z.~Chen, X.~Chen, W.~Ni, and Y.~Gao, ``{HASFL: Heterogeneity-aware Split Federated Learning over Edge Computing Systems},'' \emph{{IEEE} Trans. Mobile Comput.}, 2026.

\bibitem{sun2025intra}
Z.~Sun, X.~Guan, Z.~Lin, Z.~Fang, X.~Cai, Z.~Chen, F.~Liu, H.~Cui, J.~Xiong, W.~Ni \emph{et~al.}, ``Intra-dp: A high performance collaborative inference system for mobile edge computing,'' \emph{arXiv preprint arXiv:2507.05829}, 2025.

\bibitem{zhang2026memmark}
H.~Zhang, X.~Mao, G.~Dong, Z.~Li, X.~Su, K.~Chen, J.~Yang, and Z.~Lin, ``Memmark: State-evolution attribution watermarking for agent long-term memory systems,'' \emph{arXiv preprint arXiv:2605.25002}, 2026.

\bibitem{wei2025optimizing}
W.~Wei, Z.~Lin, X.~Liu, H.~Du, D.~Niyato, and X.~Chen, ``Optimizing split federated learning with unstable client participation,'' \emph{arXiv preprint arXiv:2509.17398}, 2025.

\bibitem{fang2024automated}
Z.~Fang, Z.~Lin, Z.~Chen, X.~Chen, Y.~Gao, and Y.~Fang, ``{Automated Federated Pipeline for Parameter-Efficient Fine-Tuning of Large Language Models},'' \emph{{IEEE} Trans. Mobile Comput.}, 2025.

\bibitem{qu2025mobile}
G.~Qu, Q.~Chen, W.~Wei, Z.~Lin, X.~Chen, and K.~Huang, ``Mobile edge intelligence for large language models: A contemporary survey,'' \emph{IEEE Communications Surveys \& Tutorials}, 2025.

\bibitem{yuan2024satsense}
H.~Yuan, Z.~Chen, Z.~Lin, J.~Peng, Z.~Fang, Y.~Zhong, Z.~Song, and Y.~Gao, ``{SatSense: Multi-Satellite Collaborative Framework for Spectrum Sensing},'' \emph{{IEEE} Trans. Cogn. Commun. Netw.}, 2025.

\bibitem{zhan2025prism}
J.~Zhan, H.~Shen, Z.~Lin, and T.~He, ``Prism: Privacy-aware routing for adaptive cloud-edge llm inference via semantic sketch collaboration,'' \emph{arXiv preprint arXiv:2511.22788}, 2025.

\bibitem{lin2026gapsl}
Z.~Lin, O.~Aouedi, W.~Ni, S.~Chatzinotas, and X.~Chen, ``Gapsl: A gradient-aligned parallel split learning on heterogeneous data,'' \emph{arXiv preprint arXiv:2603.18540}, 2026.

\bibitem{duan2025llm}
T.~Duan, Z.~Zhang, Z.~Lin, S.~Guo, X.~Guan, G.~Wu, Z.~Fang, H.~Meng, X.~Du, J.-Z. Zhou \emph{et~al.}, ``{LLM-Driven Stationarity-Aware Expert Demonstrations for Multi-Agent Reinforcement Learning in Mobile Systems},'' \emph{arXiv preprint arXiv:2511.19368}, 2025.

\bibitem{du2025dp}
X.~Du, J.~Zhu, J.~Zhou, C.-m. Pun, Z.~Lin, C.~Wu, Z.~Chen, and J.~Luo, ``Dp-trae: A dual-phase merging transferable reversible adversarial example for image privacy protection,'' \emph{IEEE Transactions on Dependable and Secure Computing}, 2025.

\bibitem{zhang2025robust}
Z.~Zhang, T.~Duan, Z.~Lin, D.~Huang, Z.~Fang, Z.~Sun, L.~Xiong, H.~Liang, H.~Cui, Y.~Cui \emph{et~al.}, ``Robust deep reinforcement learning in robotics via adaptive gradient-masked adversarial attacks,'' \emph{Proc. IROS}, 2025.

\bibitem{zhuang2026physically}
W.~Zhuang, W.~Xie, Q.~Zhang, X.~Du, Z.~Lin, Z.~Lin, H.~Cai, J.~Zhou, Z.~Fang, C.-m. Pun \emph{et~al.}, ``Physically-induced atmospheric adversarial perturbations: Enhancing transferability and robustness in remote sensing image classification,'' \emph{arXiv preprint arXiv:2604.14643}, 2026.

\bibitem{zhu2026io}
J.~Zhu, X.~Du, X.~Liu, J.-Z. Zhou, Q.~Xu, Z.~Lin, and C.-M. Pun, ``Io-rae: Information-obfuscation reversible adversarial example for audio privacy protection,'' in \emph{Proceedings of the AAAI Conference on Artificial Intelligence}, vol.~40, no.~2, 2026, pp. 1632--1640.

\bibitem{r1}
H.~Song, B.~Gu, K.~Son, and W.~Choi, ``Joint optimization of edge computing server deployment and user offloading associations in wireless edge network via a genetic algorithm,'' \emph{IEEE Transactions on Network Science and Engineering}, vol.~9, no.~4, pp. 2535--2548, 2022.

\bibitem{r3}
Z.~Liu, Q.~Z. Sheng, X.~Xu, D.~Chu, and W.~E. Zhang, ``Context-aware and adaptive qos prediction for mobile edge computing services,'' \emph{IEEE Transactions on Services Computing}, vol.~15, no.~1, pp. 400--413, 2022.

\bibitem{ra3}
X.~Yang, J.~Feng, L.~Liu, and Q.~Pei, ``Optimizing resource utilization in consumer electronics networks through an enhanced grey wolf optimization algorithm with uav collaboration,'' \emph{IEEE Transactions on Consumer Electronics}, vol.~71, no.~2, pp. 7376--7386, 2025.

\bibitem{r4}
M.~Zhang, X.~Shen, J.~Cao, Z.~Cui, and S.~Jiang, ``Edgeshard: Efficient llm inference via collaborative edge computing,'' \emph{IEEE Internet of Things Journal}, vol.~12, no.~10, pp. 13\,119--13\,131, 2025.

\bibitem{yang2025neuroagentself}
J.~Yang, V.~Govindarajan, M.~Y.~H. Al-Shamri, H.~Aldossary, A.~Ksibi, Z.~A. Shaikh, L.~Y. Por, and K.~Qi, ``Neuroagent-x: A self-evolving cognitive agent for securing consumer iot systems against ai-enabled anomalies and adversarial threats,'' \emph{IEEE Transactions on Consumer Electronics}, vol.~71, no.~4, pp. 12\,226--12\,235, 2025.

\bibitem{r5}
J.~Feng, L.~Liu, X.~Hou, Q.~Pei, and C.~Wu, ``Qoe fairness resource allocation in digital twin-enabled wireless virtual reality systems,'' \emph{IEEE Journal on Selected Areas in Communications}, vol.~41, no.~11, pp. 3355--3368, 2023.

\bibitem{ra5}
X.~Yang \emph{et~al.}, ``Irs-assisted hyperspectral image processing in satellite edge computing services,'' \emph{IEEE Transactions on Services Computing}, 2025, accepted.

\bibitem{yang2026qugridquantum}
J.~Yang, L.~Fang, Z.~Liu, Z.~A. Shaikh, A.~Ksibi, Z.~Song, S.~Prajapat, and L.~Y. Por, ``Qugrid-fed: Quantum-assisted federated iov energy and security management for sustainable urban transportation systems,'' \emph{IEEE Communications Magazine}, pp. 1--5, 2026.

\bibitem{r6}
J.~Yang, Y.~Wu, Y.~Yuan, H.~Xue, S.~Bourouis, M.~Abdel-Salam, S.~Prajapat, and L.~Y. Por, ``Llm-ae-mp: Web attack detection using a large language model with autoencoder and multilayer perceptron,'' \emph{Expert Systems with Applications}, vol. 274, 2025.

\bibitem{r7}
T.-H. Vu, S.~K. Jagatheesaperumal, M.-D. Nguyen, N.~V. Huynh, S.~Kim, and Q.-V. Pham, ``Applications of generative ai (gai) for mobile and wireless networking: A survey,'' \emph{IEEE Internet of Things Journal}, vol.~12, no.~2, pp. 1266--1290, 2025.

\bibitem{r9}
G.~Xie \emph{et~al.}, ``Gai-iov: Bridging generative ai and vehicular networks for ubiquitous edge intelligence,'' \emph{IEEE Transactions on Wireless Communications}, vol.~23, no.~10, pp. 12\,799--12\,814, 2024.

\bibitem{r10}
A.~Ucar, M.~Karakose, and N.~Kırımça, ``Artificial intelligence for predictive maintenance applications: Key components, trustworthiness, and future trends,'' \emph{Applied Sciences}, vol.~14, no.~2, p. 898, 2024.

\bibitem{r100}
J.~Yang \emph{et~al.}, ``Llmedgesec: Llm-enabled log reasoning for zero-day iot threat detection,'' \emph{IEEE Internet of Things Journal}, 2026, accepted.

\bibitem{r11}
Z.~Wu, H.~Zhang, L.~Li, Y.~Lu, and J.~Yang, ``Gai-based resource management in ris-aided next-generation network and communication,'' \emph{IEEE Transactions on Cognitive Communications and Networking}, vol.~11, no.~2, pp. 847--857, 2025.

\bibitem{r12}
C.~Zhang \emph{et~al.}, ``Gai-based resource and qoe aware service placement in next-generation multi-domain iot networks,'' \emph{IEEE Transactions on Cognitive Communications and Networking}, vol.~11, no.~2, pp. 873--885, 2025.

\bibitem{r13}
P.~Arthurs, L.~Gillam, P.~Krause, N.~Wang, K.~Halder, and A.~Mouzakitis, ``A taxonomy and survey of edge cloud computing for intelligent transportation systems and connected vehicles,'' \emph{IEEE Transactions on Intelligent Transportation Systems}, vol.~23, no.~7, pp. 6206--6221, 2022.

\bibitem{r14}
L.~Zhao \emph{et~al.}, ``A digital twin-assisted intelligent partial offloading approach for vehicular edge computing,'' \emph{IEEE Journal on Selected Areas in Communications}, vol.~41, no.~11, pp. 3386--3400, 2023.

\bibitem{r15}
H.~Liu, W.~Huang, D.~I. Kim, S.~Sun, Y.~Zeng, and S.~Feng, ``Towards efficient task offloading with dependency guarantees in vehicular edge networks through distributed deep reinforcement learning,'' \emph{IEEE Transactions on Vehicular Technology}, vol.~73, no.~9, pp. 13\,665--13\,681, 2024.

\bibitem{r16}
X.~Xiao, X.~Wang, and W.~Lin, ``Joint aoi-aware uavs trajectory planning and data collection in uav-based iot systems: A deep reinforcement learning approach,'' \emph{IEEE Transactions on Consumer Electronics}, vol.~70, no.~4, pp. 6484--6495, 2024.

\bibitem{ra1}
Y.~Wang, C.~Zhang, T.~Ge, and M.~Pan, ``Computation offloading via multi-agent deep reinforcement learning in aerial hierarchical edge computing systems,'' \emph{IEEE Transactions on Network Science and Engineering}, vol.~11, no.~6, pp. 5253--5266, 2024.

\bibitem{r18}
B.~Chen, H.~Zhou, J.~Yao, and H.~Guan, ``Reserve: An energy-efficient edge cloud architecture for intelligent multi-uav,'' \emph{IEEE Transactions on Services Computing}, vol.~15, no.~2, pp. 819--832, 2022.

\bibitem{r19}
S.~Goudarzi, S.~A. Soleymani, M.~H. Anisi, A.~Jindal, and P.~Xiao, ``Optimizing uav-assisted vehicular edge computing with age of information: An sac-based solution,'' \emph{IEEE Internet of Things Journal}, vol.~12, no.~5, pp. 4555--4569, 2025.

\bibitem{r20}
Y.~Zhao \emph{et~al.}, ``Joint content caching, service placement, and task offloading in uav-enabled mobile edge computing networks,'' \emph{IEEE Journal on Selected Areas in Communications}, vol.~43, no.~1, pp. 51--63, 2025.

\bibitem{r21}
L.~Tan, S.~Guo, P.~Zhou, Z.~Kuang, S.~Long, and Z.~Li, ``Multi-uav-enabled collaborative edge computing: Deployment, offloading and resource optimization,'' \emph{IEEE Transactions on Intelligent Transportation Systems}, vol.~25, no.~11, pp. 18\,305--18\,320, 2024.

\bibitem{r22}
G.~Sun \emph{et~al.}, ``Multi-objective optimization for multi-uav-assisted mobile edge computing,'' \emph{IEEE Transactions on Mobile Computing}, vol.~23, no.~12, pp. 14\,803--14\,820, 2024.

\bibitem{r23}
J.~Geng, Z.~Qin, and S.~Jin, ``Dynamic resource allocation for cloud-edge collaboration offloading in vec networks with diverse tasks,'' \emph{IEEE Transactions on Intelligent Transportation Systems}, vol.~25, no.~12, pp. 21\,235--21\,251, 2024.

\bibitem{r24}
J.~S. Reward, L.~T. Yang, and J.~H. Park, ``Digital twin-assisted resource allocation framework based on edge collaboration for vehicular edge computing,'' \emph{Future Generation Computer Systems}, vol. 150, pp. 243--254, 2024.

\bibitem{r25}
X.~Dai, Z.~Xiao, H.~Jiang, and J.~C.~S. Lui, ``Uav-assisted task offloading in vehicular edge computing networks,'' \emph{IEEE Transactions on Mobile Computing}, vol.~23, no.~4, pp. 2520--2534, 2024.

\bibitem{r27}
X.~Zhang \emph{et~al.}, ``Beyond the cloud: Edge inference for generative large language models in wireless networks,'' \emph{IEEE Transactions on Wireless Communications}, vol.~24, no.~1, pp. 643--658, 2025.

\bibitem{zhang2026largelanguage}
S.~Zhang, J.~Yang, Z.~Song, Z.~Lin, S.~Prajapat, Z.~Xia, H.~Ghayvat, H.~Ding, L.~Y. Por, and A.~K. Das, ``Large language model enhanced differentiable trajectory planning for iot-enabled autonomous driving,'' \emph{IEEE Internet of Things Journal}, pp. 1--1, 2026.

\bibitem{song2025smartcity}
Z.~Song, H.~Ding, L.~Jamel, J.~Yang, M.~A. Khan, J.~M. Gorriz, J.~Baili, and L.~Y. Por, ``Smart-city spatiotemporal data-driven trajectory prediction for autonomous vehicles via attention mechanisms and self-supervised learning,'' \emph{IEEE Transactions on Consumer Electronics}, vol.~71, no.~4, pp. 11\,834--11\,845, 2025.

\bibitem{r28}
X.~Dai, S.~Tian, H.~Liu, Z.~Li, H.~Jiang, and Q.~Deng, ``Joint optimization of offloading and caching in full-duplex-enabled edge computing networks,'' \emph{IEEE Transactions on Mobile Computing}, vol.~24, no.~8, pp. 6996--7011, 2025.

\bibitem{r29}
C.~Li, K.~Jiang, Z.~Zhang, C.~Xiong, and S.~Wan, ``Joint service caching and computation offloading scheme with 3d uav deployment for icvs in uav-assisted vec,'' \emph{IEEE Transactions on Communications}, vol.~73, no.~11, pp. 10\,886--10\,899, 2025.

\end{thebibliography}

\vfill

\end{document}